\documentclass[natbib,twocolumn]{svjour3} 
\usepackage{multicol}
\usepackage{graphicx}
\usepackage{amsmath}
\usepackage{amssymb}
\usepackage{natbib}
\usepackage{lipsum}
\usepackage{lineno}
\usepackage{subfigure}
\usepackage{xcolor}
\usepackage{multirow}
\usepackage{rotating}
\usepackage{tabularx}
\usepackage{lscape} 
\begin{document}


\title{An unified material interpolation for topology optimization of multi-materials}

\author{  Bing Yi  \and
       Gil Ho Yoon  \and 
       Ran Zheng \and
       Long Liu \and
       Daping Li  \and 
       Xiang Peng  %etc.
}


\institute{Bing Yi \at
             School of Traffic and Transportation Engineering, Central South University, Changsha, People's Republic of China \\
              \email{bingyi@csu.edu.cn} 
              \and
              Gil Ho Yoon  \at
              School of Mechanical Engineering, Hanyang University, Seoul, South Korea\\
              \email{ghy@hanyang.ac.kr or gilho.yoon@gmail.com} 
               \and
             Ran Zheng \at
             School of Traffic and Transportation Engineering, Central South University, Changsha, People's Republic of China 
               \and
             Long Liu \at
               School of Traffic and Transportation Engineering, Central South University, Changsha, People's Republic of China 
               \and 
              Daping Li  \at
              Institute of Engineering Innovation, China Construction Fifth Engineering Division Corp.,ltd, Changsha, People's Republic of China
               \and
              Xiang Peng  \at
              College of Mechanical Engineering, Zhejiang University of Technology, Hangzhou, People's Republic of China
}

\date{Received: date / Accepted: date}
%%%%%%%%%%%%%%%%%%%%%%%%

\maketitle

\begin{abstract}

Topology optimization is one of the engineering tools for finding efficient design. For the material interpolation scheme, it is usual to employ the SIMP (Solid Isotropic Material with Penalization) or the homogenization based interpolation function for the parameterization of the material properties with respect to the design variables assigned to each finite element. For topology optimization with single material design, i.e., solid or void, the parameterization with 1 for solid and 0 for void becomes relatively straight forward using a polynomial function. For the case of multiple materials, some issues of the equality modeling of each material and the clear 0, 1 result of each element for the topology optimization issues become serious because of the curse of the dimension. To relieve these issues, this research proposes a new mapping based interpolation function for multi-material topology optimization. Unlike the polynomial based interpolation, this new interpolation is formulated by the ratio of the $p$-norm of the design variables to the 1-norm of the design variable multiplied by the design variable for a specific material. With this alternative mapping based interpolation function, each material are equally modeled and the clear 0, 1 result of each material for the multi-material topology optimization model can be improved. This paper solves several topology optimization problems to prove the validity of the present interpolation function.

\end{abstract}
%%%%%%%%%%%%%%%%%%%%%%%%
\section{Introduction}
\label{sec1}

 Often components, modules and systems are usually made with multiple materials and multiple components. Different components manufactured for structures or products are assembled together rather than one single manufacturing process. Recently the development of 3D printing or additive manufacturing technology even makes it enable to manufacture components with multi-materials in a single machine without assembling \cite{Vidimce2013,Gaynor2014,BANDYOPADHYAY20181}. As the performance can be significantly affected by the  locations of multiple materials, it is important to optimize the distributions of multi-materials and the topology layout of multiple materials. 

\begin{figure}
\centering
\includegraphics[scale=0.2]{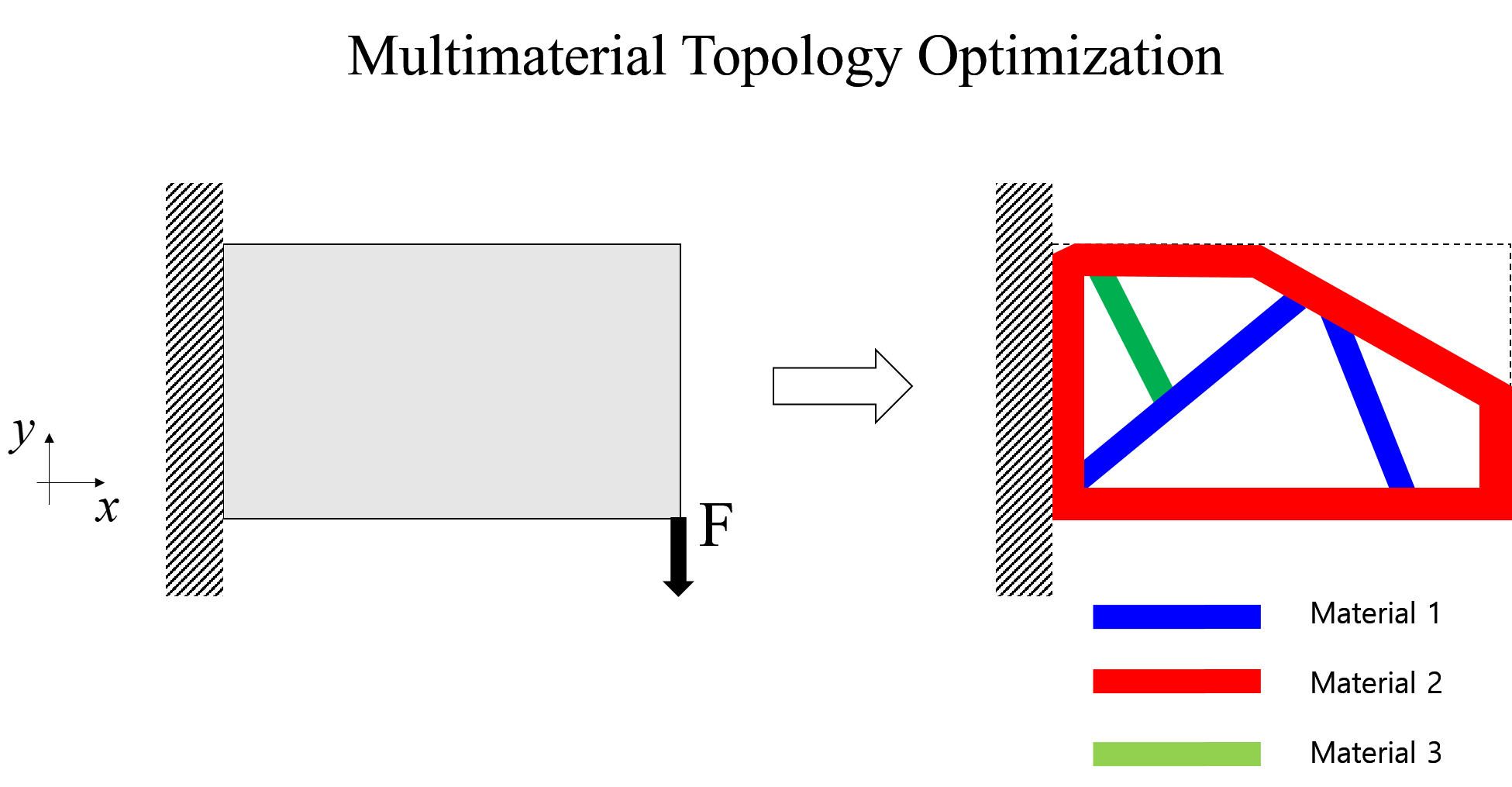} 
    \caption{An illustration of topology optimization with three materials; various interpolation functions have been proposed.} 
    \label{fig:multimaterialTO}
\end{figure}

Topology optimization is one of the standard methods for multiple materials design and optimization as shown in Fig.~\ref{fig:multimaterialTO}. There are vast researches regarding the topology optimization for multiple materials. The purpose of this topology optimization problem is to find out the optimal material distributions with multiple materials (multiple phases) for structural problem and multi-phyiscs problems. With the help of the development of mathematical optimization theory and advanced computational scheme, the size, shape and topology optimization schemes have been widely adopted and developed for scientists and engineers \citep{BENDSOE1988197,Sigmund2013,WANG2003227,ALLAIRE2004363,XIE1993885,Peng2017,Yoon2019,YI2019593,YiEo2021,YiIjnme2021}. Among them, this research focuses on the topology optimization method providing an optimal layouts for complex engineering structures without a prior given optimal topology. The demand towards multiple materials and multiple components is on the rise. Many introductory materials for structural optimization problem can be found and it becomes easier to conduct researches for structure problem. In the framework of the topology optimization method, the design variables assigned to each finite element interpolate the material properties of the physical equations of interest to find out void and material statuses. By extending the single phase material design problem into multiple phases, lot of methods were proposed to solve the problem for topology optimization of multi-materials. 

We summarized the methods for topology optimization of multi-materials into three groups. One of the widely accepted methods for topology optimization of multi-materials is the SIMP or extended SIMP method, which interpolate the involved material properties with the simple polynomial function. They can be modeled with single variable or multiple variables for multi-materials interpolation. Multiple variables based modeling method has the advantage of clearly definition of each component. \cite{Bendsoe1999,Bendsoe2003} first extended the SIMP interpolation for multi-material by combining exponential function. \cite{Tavakoli2014,PARK2015571} followed the work to easily implement it into traditional optimizer by using alternating active-phase algorithm, which is efficiently for the employing of an additional outlier optimization iteration. A simple modification with linear combination of penalty interpolation function for multi-material were used for topology optimization of laminated composite beam cross sections by \cite{BLASQUES20123278,BLASQUES201445}. \cite{Cui2018} modified the density interpolation approach based on the logistic function, which eﬀectively realizes the polarization of the intermediate-density elements. \cite{Long2018} introduced the reciprocal variables into the  the formulation of topology optimization to overcome these undesirable local optimum phenomena for multi-material optimization. \cite{Bohrer2021} extended it for multi‑material based micro-structural topology
optimization of the functionally graded materials and \cite{Bohrer2021} used it for multi-material topology optimization considering both isotropic and anisotropic materials. A similar method, called  RAMP material interpolation schemes is also involved into the optimization of multi-material \cite{Hvejsel2011,Dzieranowski2012,WU20191136}. However, these methods rely on a large number of sparse linear constraints to enforce the selection of at most one material in each design subdomain. \cite{TAVAKOLI2014534} proposed a new objective function for multi-material optimization by introducing a Ginzburg–Landau energy term. It firstly solves the optimization on the L2 space by the projected steepest descent algorithm, and then project the parameters onto the feasible domain with a practically time-linear algorithm. 

There is a simple interpolation function for multi-material optimization called discrete material optimization was first adopted for multi-material by \cite{Gao2011}, which was proposed by \cite{DMO2004} for topology optimization of composite structure. \cite{Bruyneel2011a,Bruyneel2011b} proposed a new interpolation approach, called shape function parameterization (SFP) for multi-material optimization to reduce the number of design variables in DMO. \cite{Kennedy2015} proposed a scalable approach for large scale topology optimization with the DMO material properties interpolation function.  \cite{SANDERS2018798,SANDERS2019} proposed a simple and robust formulation for multi-materials topology optimization by taking advantage of  the separable dual objective in the linearized sub-problems, the ZPR scheme proposed by \cite{Zhang2018} was used for updating volume/mass constraint independently. Hence, the formulation is effective for multiple volume/mass constraints that can control all or a subset of the candidate materials in the entire domain or a subset of the domain. \cite{NGUYEN201979} extended it for the multi-material topology optimization by using polytree-based adaptive polygonal finite elements. 

Some researchers also tried to use single variable to model the topology optimization of multi-materials. \cite{Yin2001} proposed the peak function based material interpolation model, which uses a linear combination of a normal distribution functions. It can be easily incorporated into the topology optimization without increasing the number of design variables. However, the horizontal zero slope of interpolation curve is a potential source of difficulty in the numerical calculations for that it is impossible to cross such a singular point to make transition from one material phase to another during the optimization process. \cite{Zuo2017} proposed the ordered SIMP interpolation function by introducing the power functions with scaling and translation coefficients  for multiple materials with respect to the normalized density variables. It indeed reduced the computation cost for optimization, but also need special interpolation function for Young's modular and cost to get a converge result.

The phase field method and the levelset method have the advantage of clear definition of material boundary, and have also been employed for the topology optimization of multi-material. \cite{Zhou2006a,Zhou2006b} introduced a general method to solve multiphase structural topology optimization problems based on Cahn–Hilliard equation. One of the most important benefits of this method is the intrinsic volume preserving property, which will be kept strictly feasible with respect to the design domain without any further effort. However, the slow convergence of the phase field method remains the main drawback of this approach. \cite{Blank2012,Blank2013} employed the volume constrained Allen–Cahn equation into the Cahn–Hilliard based topology optimization methods to overcome the slow convergence. \cite{WANG2004469} extended the single levelset method to multiple levelset, called color level sets, for structural topology optimization with multiple materials. It only need $\log_2 m$ levelset for $m$ material design, and then the structure update by using a set of Hamilton-Jacobi equation. \cite{WANG20151570,LIU2016113} followed the work by using $m$ level set functions to represent $m$ materials and one void phase (totally M+1 phases). They claimed for the advantage of guaranteeing that each point contains exactly one phase without overlaps and with an explicit mathematical expression, which greatly facilitates the design sensitivity analysis. \cite{GUO2014632,Chu2018} extended multiple levelset modeling method for the topology optimization of multi-material with stress constraint. \cite{CUI201641} proposed a level-set based multi-material topology optimization method using a reaction diffusion equation. It modified the multi-material description from Multi-Material Level Set (MM-LS)  proposed by \cite{WANG20151570}, which also has the advantage of that each phase is represented by a combined formulation of different level set functions. \cite{LIU2018444} proposed a new multi-material level set topology optimization method by using each level set function to represent one material phase and the overlapping areas are filled with an artificial material type. It has the advantage of keeping the signed distance information with individual material regions, and can successfully realize the component length scale controlling of multi-material structures. \cite{Wang2021} extended it to the multi‑material topology optimization method by involving the material‑field series‑expansion model to reduce the number of design variables. Though all these methods perform well for topology optimization of multi-materials, the final topology highly depending on the initial design remains one big issue to be solved especially for multi-materials. 

There are also a few researches are introduced to improve the efficiency for multi-materials optimization by using the discrete variable based topology optimization. \cite{Ramani2010,Ramani2011}derived the pseudo sensitivity for the discrete material variable and constructed the heuristic scheme to obtain the solution. \cite{YANG2018182} used the elemental compliance to update the density during the iteration of optimization procedure, and also a practical regulated iterative numerical approach was involved to find the solution to the multi-material topology optimization problem by solving a series of two-material sub-problems. Although, these methods improved the efficiency for multi-material optimization, but the pseudo sensitivity decreased accuracy and robustness for topology optimization problem with a large number of design variables.

\begin{figure}
\centering
    \subfigure[]{
     \centering
        \includegraphics[scale=0.2]{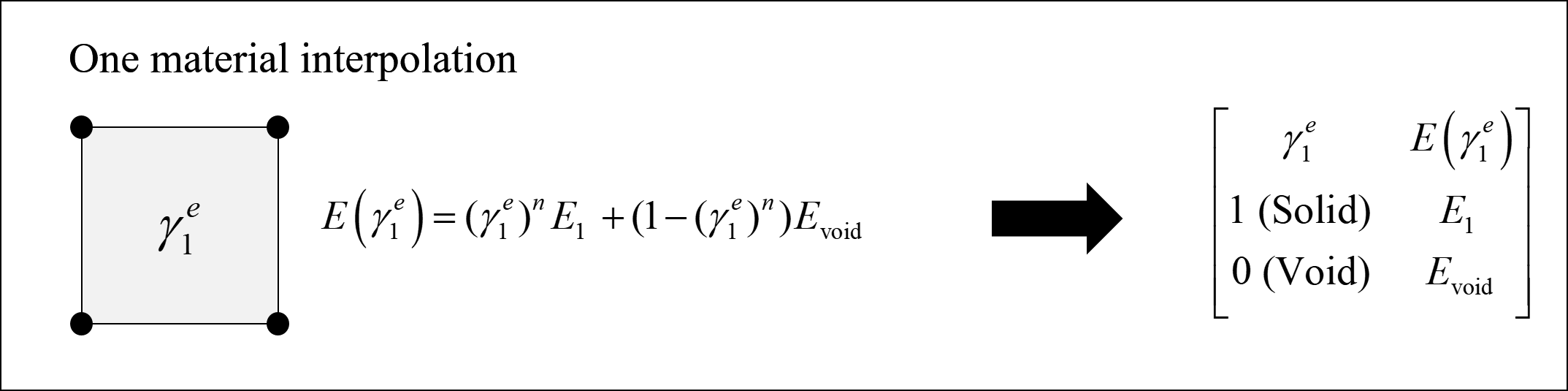}
          \label{(a)}} 
    \subfigure[]{
     \centering
        \includegraphics[scale=0.2]{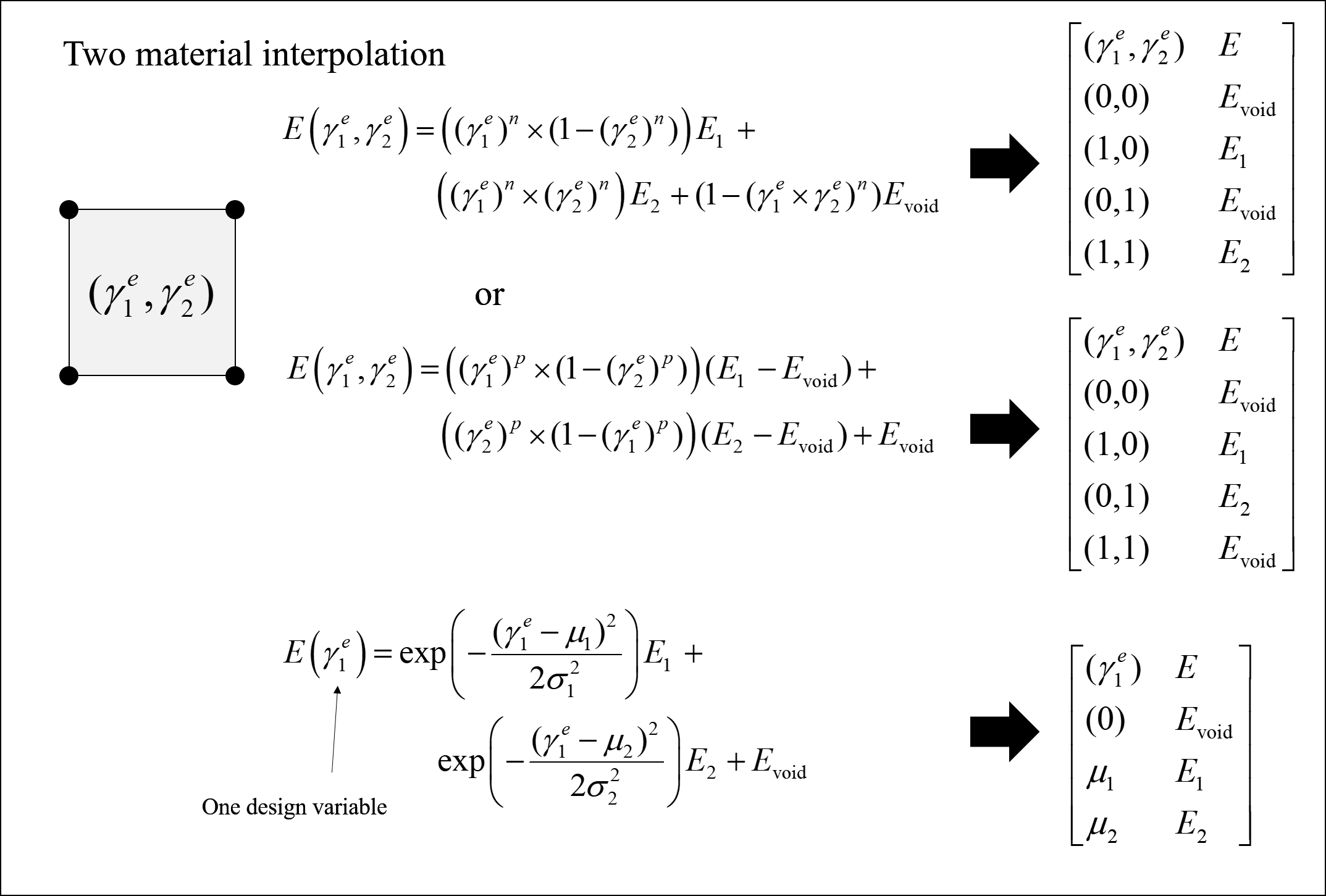}
          \label{b)}}
         
    \caption{Conventional interpolation schemes. (a) The interpolation function (SIMP) for one material and (b) the interpolation functions for two materials.}
    \label{fig:Conventionalinterpolationfunction}
\end{figure}

For multiple materials of SIMP based topology optimization, one of the most important things is the interpolation function for Young’s modular. The main purpose of the interpolation function in topology optimization is to relax the discrete optimization problem with the continuous design variables. In other words, the original topology optimization problem is the integer optimization problem with zeros and ones. To solve the integer hard problem, it is relaxed with the continuous design variables. Then local optima with discrete void and solid are pursued after optimization. In these procedures, it becomes important to employ a proper relaxation function interpolating material properties with the continuous design variables to obtain designs with limit values. With an improper interpolation function, optimal layouts with a lot of grey design variables are prone to be generated. This issue becomes serious even with one-material design, i.e., void or solid. For example, the SIMP method with a smaller penalization leads many grey elements. This difficulty becomes more serious with multiple materials. This difficulty lies in the fact that the local optimum issue becomes serious in case of multiple materials. Some relevant researches have proved that the extension of the SIMP method can be applied for the interpolation function of multiple materials and its extension have been applied for many layout optimization problems. The extension of the single material method is obvious and it is also possible to find the research adopting the combined exponential functions for multiple materials as shown in \ref{fig:Conventionalinterpolationfunction}.

The revisiting of the interpolation functions for topology optimization with multiple materials in connection to multiple components is necessary. It is challenging as through the optimization, the distinct materials and design variables should be appeared to satisfy the volume constraints and overcome the non-convex interpolation function. Hence, the issue of the interpolation function is one of the key aspects of topology optimization for multiple materials. One of the difficulties faced with this is the selecting of the proper and good initial design variables and not pushing the optimizer to find out a serious local optimum. To relieve this difficulty and contribute this research subject, it is necessary to revisit the subject of the interpolation function for multiple materials as it directly relates to the clear 0, 1 result of the topology optimization method. 

To our best knowledge, there have been several researches about the interpolation itself based on polynomial functions. Most of the researches are based on the polynomial SIMP based interpolation function and it extends the scopes of its application of multiple phases and multi-layered structures. The present research aims to contribute this subject too. The present study proposes to use the mapping based interpolation function for topology optimization for multiple materials. Rather than the polynomial interpolation function or the SIMP based interpolation, this mapping based interpolation function combines the $p$-norm of the design variables assigned to each finite element and the 1-norm of the design variable. Compared with the polynomial based interpolation function, this mapping based interpolation has the advantages in case of plentiful materials. To prove the concept of the present mapping based interpolation function, several topology optimization examples are solved.  

This paper is organized as follows. Section 2 describes the basic equations for the structural optimization problem and develops a new mapping based interpolation function for multiple materials and discusses its characteristics to the existing interpolation function. Section 3 presents several numerical examples to show the advantages and disadvantages of the mapping based interpolation function. Section 4 presents the conclusions and suggests future research topics. 

%\input{test}
%%%%%%%%%%%%%%%%%%%%%%%%
\section{A new unified mapping based interpolation function}
\label{sec2}

Before presenting the new mapping based interpolation function, this section explains the topology optimization problem and the SIMP based interpolation functions for multiple materials. 

\subsection{Linear elasticity equation and topology optimization}

To develop an unified mapping based interpolation function, the topology optimization problem of compliance minimizing subject to volume constraint is considered here. The governing equation to solve the equilibrium equation on the domain  $\Omega $ is generally formulated as follows:

\begin{equation}
\nabla \cdot \,{\boldsymbol{\mathbf{\sigma }}}\left( {\mathbf{u}} \right) + {\mathbf{b}} = \mathbf{0}{\text{ in }}\,{\Omega } 
\end{equation}
where the Cauchy stress tensor, the displacement  vector, and the body force are denoted by ${\boldsymbol{\mathbf{\sigma }}}$, ${\mathbf{u}}$ , and  ${\mathbf{b}}$, respectively. The Dirichlet boundary condition along $\partial {\Omega _D}$ and the Neumann boundary condition along $\partial {\Omega _N}$ are defined as follows:

\begin{equation}
\begin{gathered}
{{\mathbf{u}} = {\mathbf{0}}{\text{ on }}\partial {\Omega_D}} \\
\boldsymbol{\sigma} \cdot {\mathbf{n}} = {\mathbf{f}}{\text{ on }} \partial{\Omega_N} 
\end{gathered} 
\end{equation}
where ${\mathbf{f}}$ and ${\mathbf{n}}$ denote the surface traction and the unit normal vector. The linear strain $\boldsymbol{\mathbf{\varepsilon}}$ and stress $\boldsymbol{\mathbf{\sigma}}$ relationship with the constitutive matrix ${\mathbf{C}}$ is assumed. 

\begin{equation}
\boldsymbol{\mathbf{\sigma }} = \boldsymbol{\mathbf{C\varepsilon}}
\end{equation}
Without the loss of generality, the finite element procedure is applied to calculate the structural displacements, it can be expressed as:
\begin{equation}
 {\mathbf{K}}{\mathbf{U}}{\text{ = }}{\mathbf{F}} 
\end{equation}
where the stiffness matrix, the displacement and the force vectors are denoted by ${\mathbf{K}}$ , ${\mathbf{U}}$ and ${\mathbf{F}}$, respectively.

With the multiple design variables assigned to each finite element, the following topology optimization problem can be formulated. 

\begin{equation}
\begin{gathered}
\label{eq:original op1}
  \mathop {{\text{Min}}}\limits_{\mathbf{\gamma }} {\text{  }}c{\text{ = }}{{\mathbf{F}}^{\text{T}}}{\mathbf{U}} \hfill \\
  {\text{Subject to  V(}}{\mathbf{ \gamma }}{\text{)}_i} \leqslant {\text{ V}_i^0}, i=1,...,NM \hfill \\
  {\mathbf{K}}{\text{(}}{\mathbf{\gamma }}{\text{)}}{\mathbf{U}}{\text{ = }}{\mathbf{F}} \hfill \\
{\mathbf{\gamma }} = \left[ {\begin{array}{*{20}{l}}
  {{\gamma _1^1}}&{{\gamma _2^1}}& \cdots &{{\gamma _{NM - 1}^1}}&{{\gamma _{NM}^{1}}} \\ 
  {{\gamma _{1}^2}}&{{\gamma _{2}^{2}}}& \cdots &{{\gamma _{NM - 1}^{2}}}&{{\gamma _{NM}^{2}}} \\ 
   \vdots & \vdots & \vdots & \vdots & \vdots  \\ 
  {{\gamma _{1}^{NE - 1}}}&{{\gamma _{2}^{NE - 1}}}& \cdots &{{\gamma _{NM - 1}^{NE - 1}}}&{{\gamma _{NM}^{NE - 1}}} \\ 
  {{\gamma _{1}^{NE}}}&{{\gamma _{2}^{NE}}}& \cdots &{{\gamma _{NM - 1}^{NE}}}&{{\gamma _{NM}^{NE}}} 
\end{array}} \right] \hfill \\ 
{0\leqslant{\gamma}\leqslant{1}} \hfill \\ 
\end{gathered}
\end{equation}

 The volume of the $i_{th}$ material and the allowed maximum volume are denoted by V$_i$ and $\text{V}_i^0$, respectively. The number of the finite elements and the number of the multiple materials are denoted by $NE$ and $NM$, respectively. The design variables are $\mathbf{\gamma }$ with the  subscript indicating the material properties and the superscript indicating the index of the finite element. It is common to employ the design variables same with the number of materials, i.e., two design variables for two materials and three design variables for three materials. Then, the stiffness matrix is assembled as follows: 

\begin{equation}
\begin{gathered}
{\mathbf{K}}{\text{(}}{\mathbf{ \boldsymbol\gamma }{\text{)}}{\mathbf{U}}{\text{ = }}{\mathbf{F}},{\mathbf{K}}{\text{(}}{\mathbf{\boldsymbol\gamma }}{\text{)}} = \sum\limits_{e = 1}^{NE} {{{\mathbf{k}}^e}} ({\phi_i(\gamma)})}, i=1,...,NM
\end{gathered}
\end{equation}

Conventionally the following extended SIMP based material interpolation functions for these weight factors have been proposed for topology optimization with multiple materials (See \cite{Tong2011,Zuo2017} and references therein).

\begin{equation}
{{{k}}^e} = \sum\limits_{i = 1}^{NM} {\phi _i ({{{E}}_i-E_\text{void}})+E_\text{void}} \\
\end{equation}

\begin{equation}
\phi_i=
(\gamma_1^e)^n \left[ {\mathop \prod \limits_{j = 2}^{i} (1 - {{(\gamma_{i+1 \ne NM+1}^e)}^n}){{(\gamma_j^e)}^n}} \right]
\end{equation}
where the interpolation functions are denoted by $\phi_i$ for the factor multiplied to the $i_{th}$ Young's modulus. On the other hand, the following DMO interpolation functions are also proposed \citep{DMO2004}.

\begin{equation}
\phi_i=
{\left[ {(\gamma_i^e)^n \mathop \prod \limits_{j = 1,j \ne i}^{NM} (1 - {{(\gamma_{j}^e)}^n})} \right]}
\end{equation}

%%%%%%%%%%%%%%%%%%%%%%%%%%%%

With the gradient based optimizer, the sensitivity values of the objective and the constraint with respect to the ${\mathbf{\gamma}_{i}^{e}}$ design variable can be computed as follows: 

\begin{equation}
\frac{\partial c}{\partial  {\mathbf{\gamma}_{i}^{e}}} =  - {\mathbf{U}^{\text{T}}}\frac{{\partial {\mathbf{K}}}}{{\partial {\mathbf{\gamma}_{i }^{e}}}}{\mathbf{U}}
\end{equation}

The optimization problem with the above gradient values can be solved with a finite element strategy using a gradient-based optimizer. For the topology optimization, it is crucial to interpolate the material properties with respect to the continuous density variable in the SIMP method to avoid the discrete 0-1(void or solid) optimization problem.

\subsection{Issues of the interpolated Young's modulus}

To converge the design variables to material phases and void, any interpolation function should make the stiffness contribution reduced for intermediate variables or penalized as Equation.(\ref{eq:interpolation explain}). In case of one material, the curve interpolating the envolved Young's modulus should be less than the function value of the volume ratio or less economically. This condition should be satisfied for multiple materials too; In addition, the convexity of the interpolation function should be considered.
 \begin{equation}
\begin{gathered}
\label{eq:interpolation explain}
\begin{array}{l}
\text{Case 1 (One material): }NM= 1 , \\ E= E_1 \phi +E_{void}, \phi =(\gamma^e)^n,  \\
\frac{E}{E_1} \simeq (\gamma^e)^n \le \gamma^e \\ \\ 
\text{Case 2  (multiple materials): }NM \ge 2 \\ 
\frac{E}{E_i} \le \text{the } i\text{-th volume interpolation function} 
\end{array}
\end{gathered}
\end{equation}
The present study puts the conventional interpolation functions in question as the values and the orders of Young's moduli are not considered in the interpolation functions in topology optimization (The second condition of Equation.(\ref{eq:interpolation explain})). 

In order to illustrate this issue, let us consider the three materials with 5 N/m$^2$, 2 N/m$^2$ and 1 N/m$^2$ for the Young's moduli. The interpolated Young's modulus may be formulated with the extended SIMP interpolation function as follows:

 \begin{equation}
\begin{gathered}
\label{eq:interpolation 1}
\begin{array}{l}
{\phi _1} = \gamma _1^n \times (1 - \gamma _2^n)\\
{\phi _2} = \gamma _1^n \times \gamma _2^n \times (1 - \gamma _3^n)\\
{\phi _3} = \gamma _1^n \times \gamma _2^n \times \gamma _3^n\\
E = {E_1}{\phi _1} + {E_2}{\phi _2} + {E_3}{\phi _3}
\end{array}
\end{gathered}
\end{equation}
With the DMO (Discrete Material Optimization) approach, the material property is interpolated as follows:
 \begin{equation}
\begin{gathered}
\label{eq:interpolation 1}
\begin{array}{l}
{\phi _1} = \gamma _1^n \times (1 - \gamma _2^n) \times (1 - \gamma _3^n)\\
{\phi _2} = \gamma _2^n \times (1 - \gamma _1^n) \times (1 - \gamma _3^n)\\
{\phi _3} = \gamma _3^n \times (1 - \gamma _1^n) \times (1 - \gamma _2^n)\\
E = {E_1}{\phi _1} + {E_2}{\phi _2} + {E_3}{\phi _3}
\end{array}
\end{gathered}
\end{equation}
The Young's moduli are denoted by $E_1$, $E_2$ and $E_3$, respectively. Note that the orders of the Young's moduli are not discussed or neglected in the above interpolation  Equation.(\ref{eq:interpolation 1}) and we can arbitrary choose the orders of the Young's moduli as follows:
 \begin{equation}
\begin{gathered}
\label{eq:interpolation 1}
\begin{array}{l}
\text{Case 1: } E = 5{\phi _1} + 2{\phi _2} + 1{\phi _3} \\
\text{Case 2: } E = 5{\phi _1} + 1{\phi _2} + 2{\phi _3} \\
\text{Case 3: } E = 2{\phi _1} + 5{\phi _2} + 1{\phi _3} \\
\text{Case 4: } E = 2{\phi _1} + 1{\phi _2} + 5{\phi _3} \\
\text{Case 5: } E = 1{\phi _1} + 5{\phi _2} + 2{\phi _3} \\
\text{Case 6: } E = 1{\phi _1} + 2{\phi _2} + 5{\phi _3} \\
\end{array}
\end{gathered}
\end{equation}

\begin{figure}
\centering
  \begin{tabular}{cc}
     \subfigure[]{ 
     \centering
        \includegraphics[width=40mm,trim=0cm 0cm 0cm 0cm,clip]{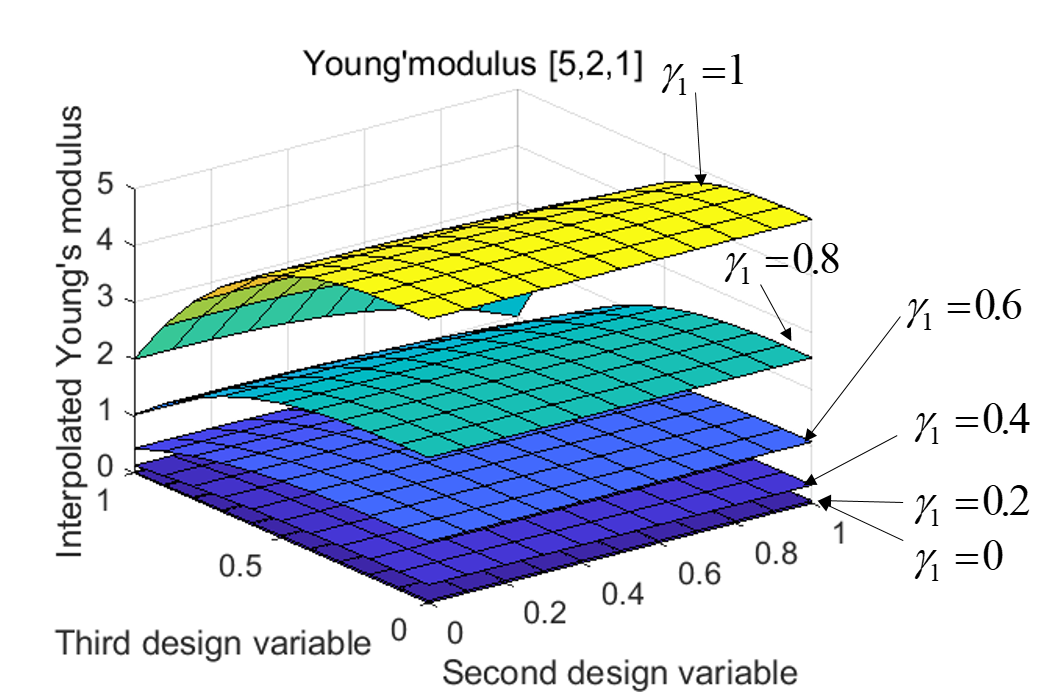}          }        
  
  \subfigure[]{
     \centering
        \includegraphics[width=40mm,trim=0cm 0cm 0cm 0cm,clip]{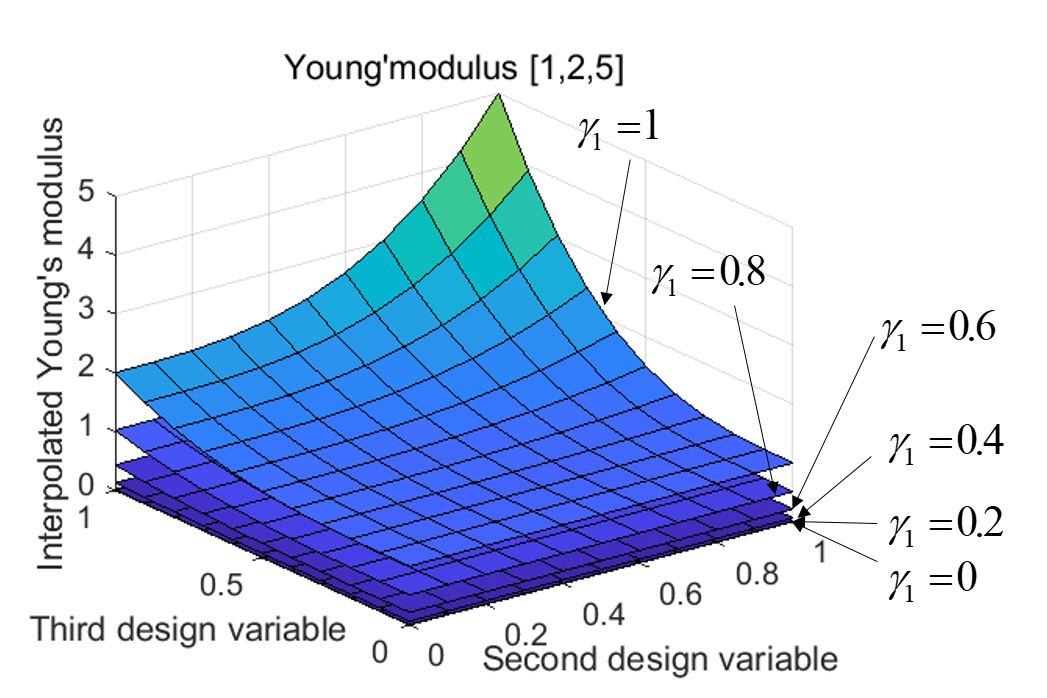}          }  \\
  \subfigure[]{
     \centering
        \includegraphics[width=40mm,trim=0cm 0cm 0cm 0cm,clip]{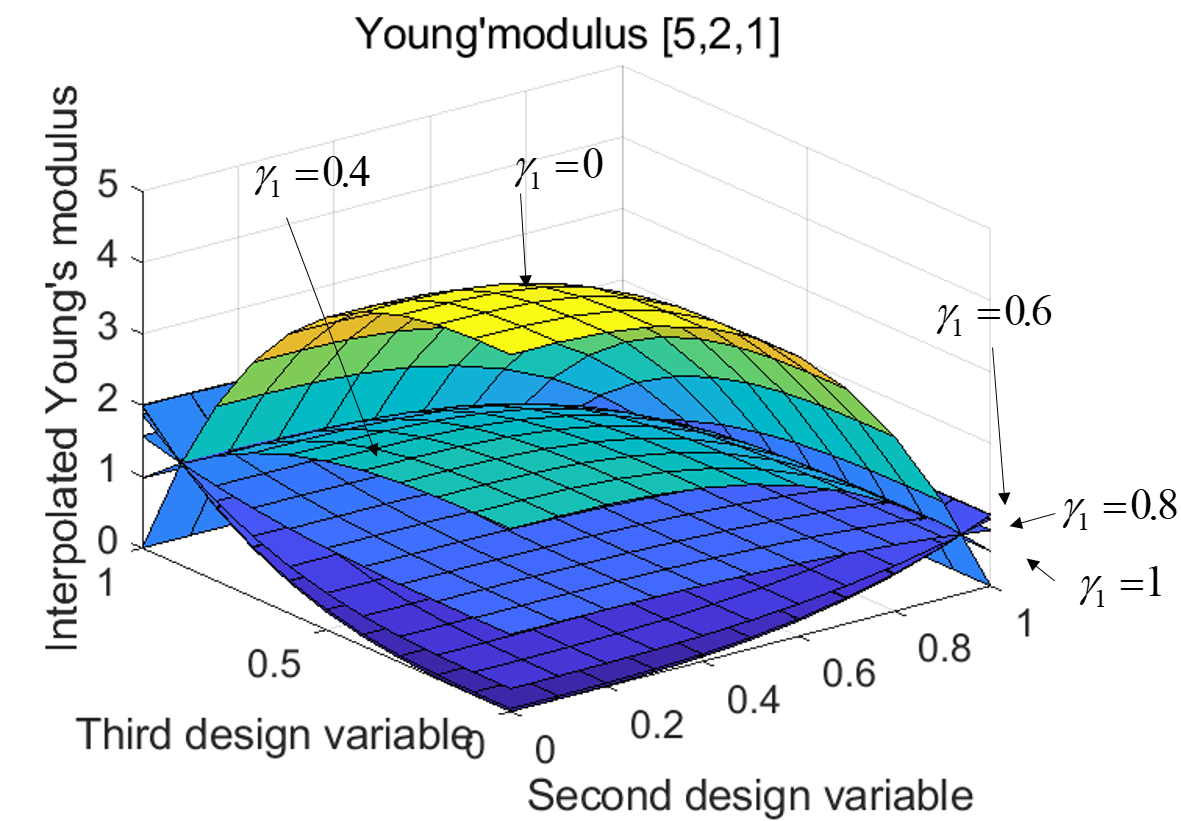}          }   
  
  \subfigure[]{
     \centering
        \includegraphics[width=40mm,trim=0cm 0cm 0cm 0cm,clip]{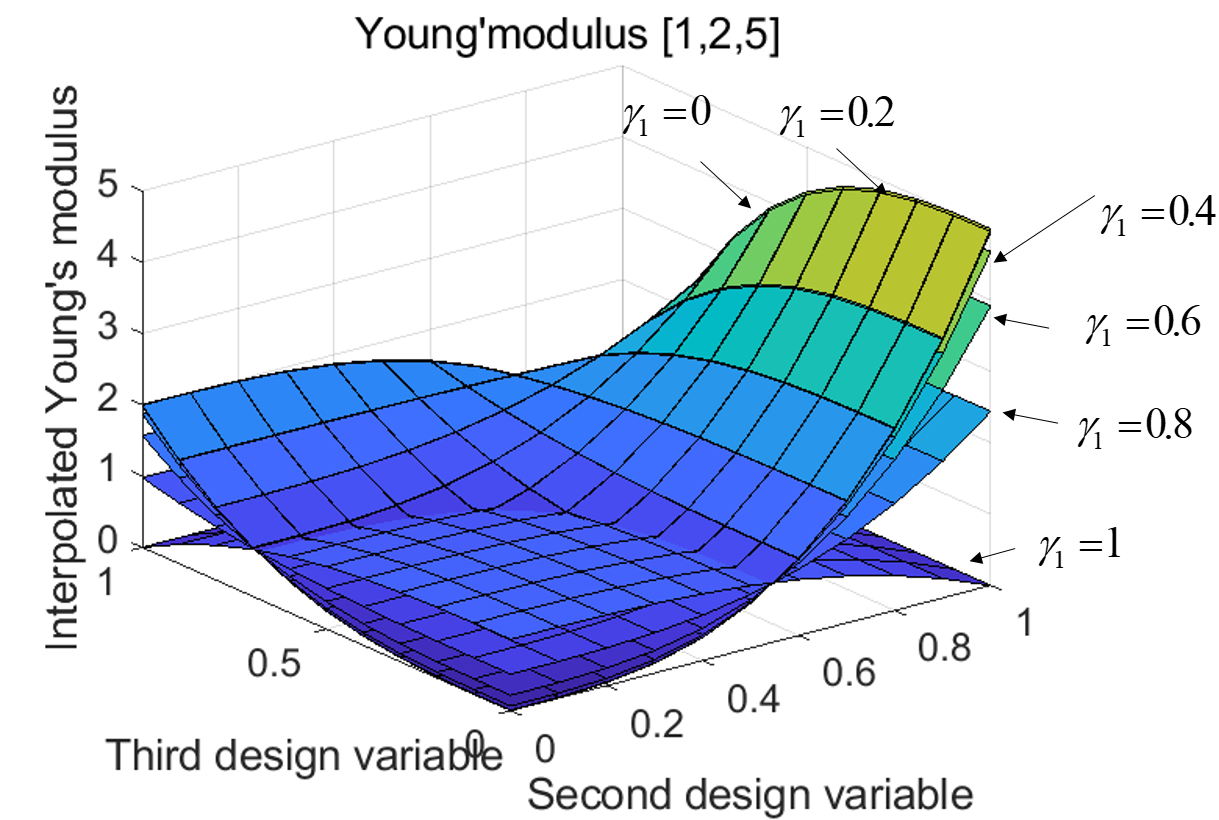}          }           
  \end{tabular}        
    \caption{Interpolated Young's modulus : (a)  the extended SIMP approach (Case 1: $n$=3, $E_1$ =5 N/m$^2$,$E_2$ =2 N/m$^2$, $E_3$ =1 N/m$^2$), (b) the extended SIMP approach (Case 6: $n$=3, $E_1$ =1 N/m$^2$,$E_2$ =2 N/m$^2$, $E_3$ =5 N/m$^2$), (c) the DMO approach  (Case 6: $n$=3, $E_1$ =1 N/m$^2$,$E_2$ =2 N/m$^2$, $E_3$ =5 N/m$^2$) and (d) the DMO approach  (Case 1: $n$=3, $E_1$ =5 N/m$^2$,$E_2$ =2 N/m$^2$, $E_3$ =1 N/m$^2$).}
    \label{fig:interpolatedYoungmodulus}
\end{figure}

To converge the design variables into ones or zeros, the convex curves would perform well with respect to the design variables for the above interpolation functions with sufficient penalization factors or the ratio of the interpolated Young's modulus to the $i$th Young's modulus should be less than each volume interpolation function. Commonly these conditions are not satisfied and was discussed in the SIMP approach, and then the ordered multi-material SIMP interpolation was proposed \citep{Zuo2017,Kato}. 
For example, Figure~\ref{fig:interpolatedYoungmodulus}  shows the distributions of the above interpolated functions (Case 1 and Case 6). {The properties of the curves depend on the design variables and the order of the Young's moduli.} The order of the material properties affects the characteristics of the surfaces and in this case, the sufficient penalizations cannot be achievable. For example, the normalized interpolated Young's modulus should be less than the volume ratio for the clear result of each element. It is not clear and often cannot be achievable for multiple materials (See Equation.~(\ref{eq:interpolation explain})).

Due to the above issues of the interpolated Young's modulus, it is found that it is difficult to guarantee the clear 0, 1 result for the design variables for multiple materials with the penalization factors and the local optima issue becomes a serious problem.

\subsection{A new unified mapping based interpolation function for multiple materials}

\begin{figure}[h]
\centering
    \subfigure[]{
    \centering
        \includegraphics[scale=0.35]{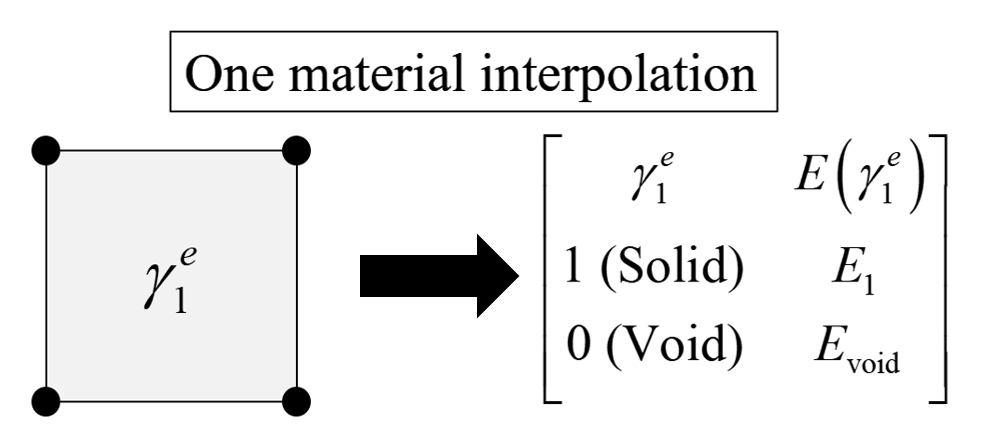}
          \label{(a)}} 
     \subfigure[]{
    \centering
\includegraphics[scale=0.35]{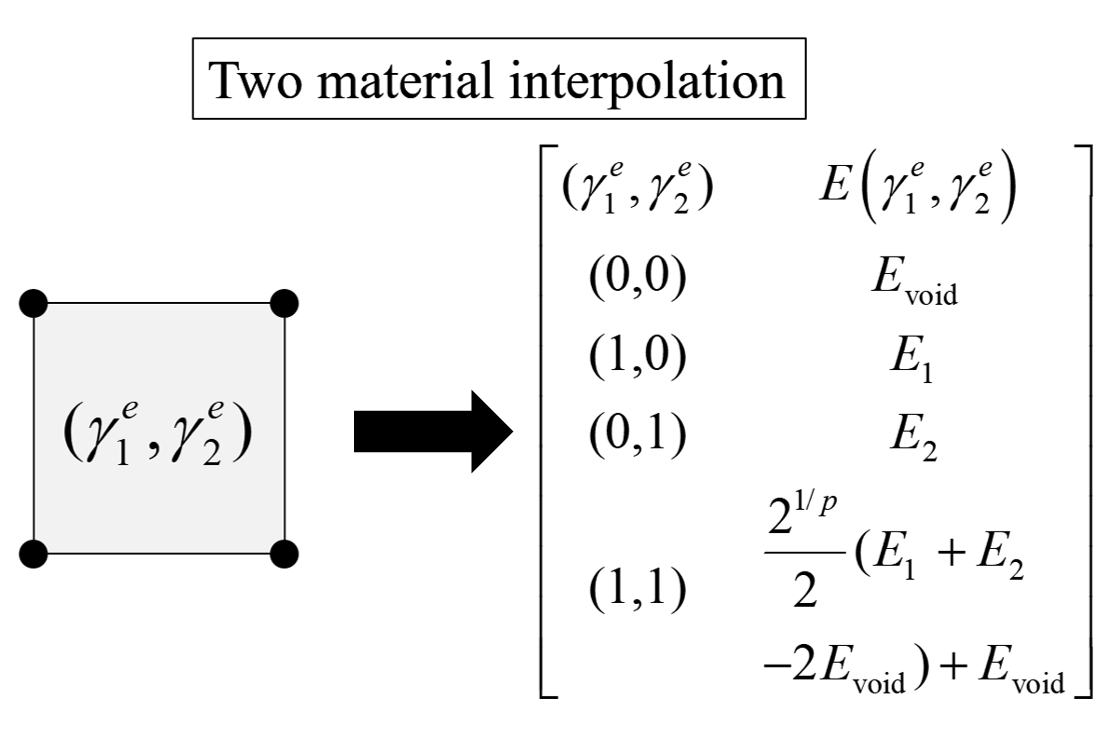}
          \label{(b)}} 
     \subfigure[]{
    \centering
\includegraphics[scale=0.35]{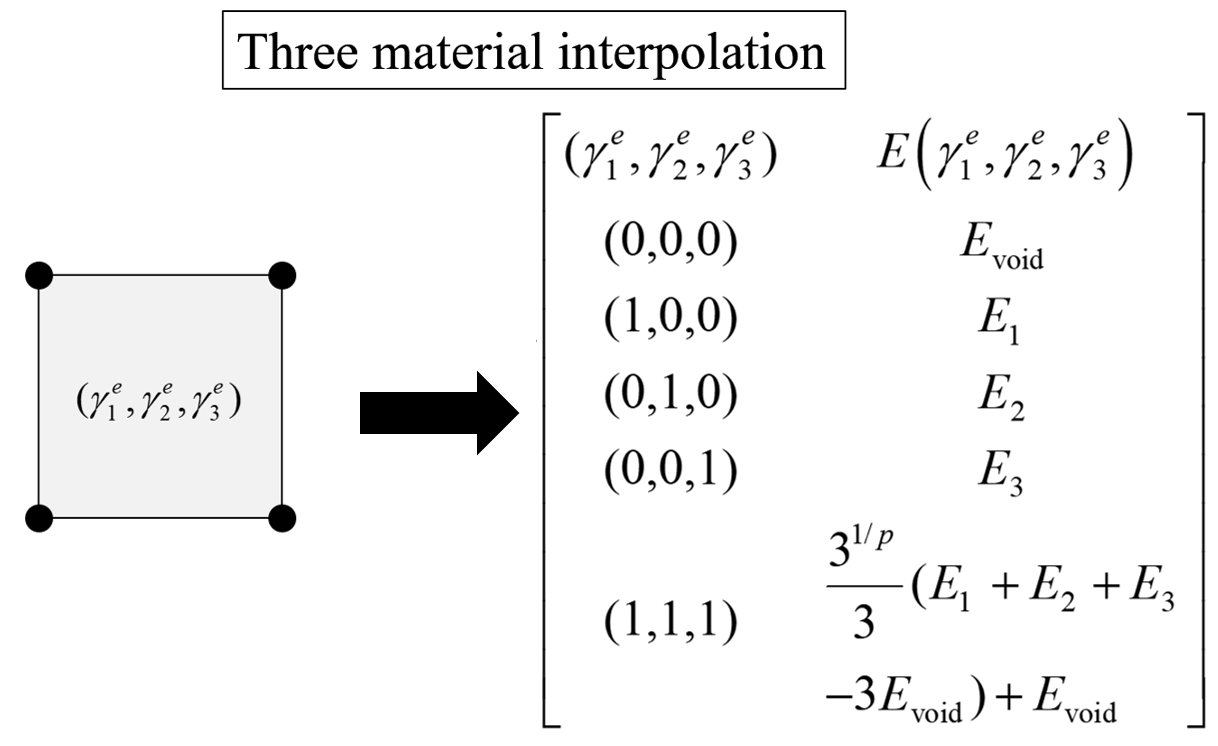}
          \label{(c)}}           
            
    \caption{A new mapping based interpolation function for multiple materials. (a) the formulation for the one material interpolation ,(b) the formulation for two materials and (c) the formulation for three materials.}
\label{fig:newmappingformulation}
\end{figure}
Bearing the characteristics of the above interpolation functions of topology optimization, this subsection proposes a new unified mapping based interpolation function illustrated in Fig. \ref{fig:newmappingformulation} for multiple materials. The present approach is also started at the finite element level and the material properties are expressed as a weighted sum of material properties of interest. Compared with the existing interpolation functions based on the polynomial series or the SIMP based interpolation functions, the present interpolation function is formulated based on the ratio of the $p$-norm of the design variables to the sum or the $1$-norm of the design variables multiplied by the design variable for the corresponding material as follows:

\begin{figure}
\centering
  \begin{tabular}{cc}
     \subfigure[]{ 
     \centering
        \includegraphics[width=80mm,trim=0cm 0cm 0cm 0cm,clip]{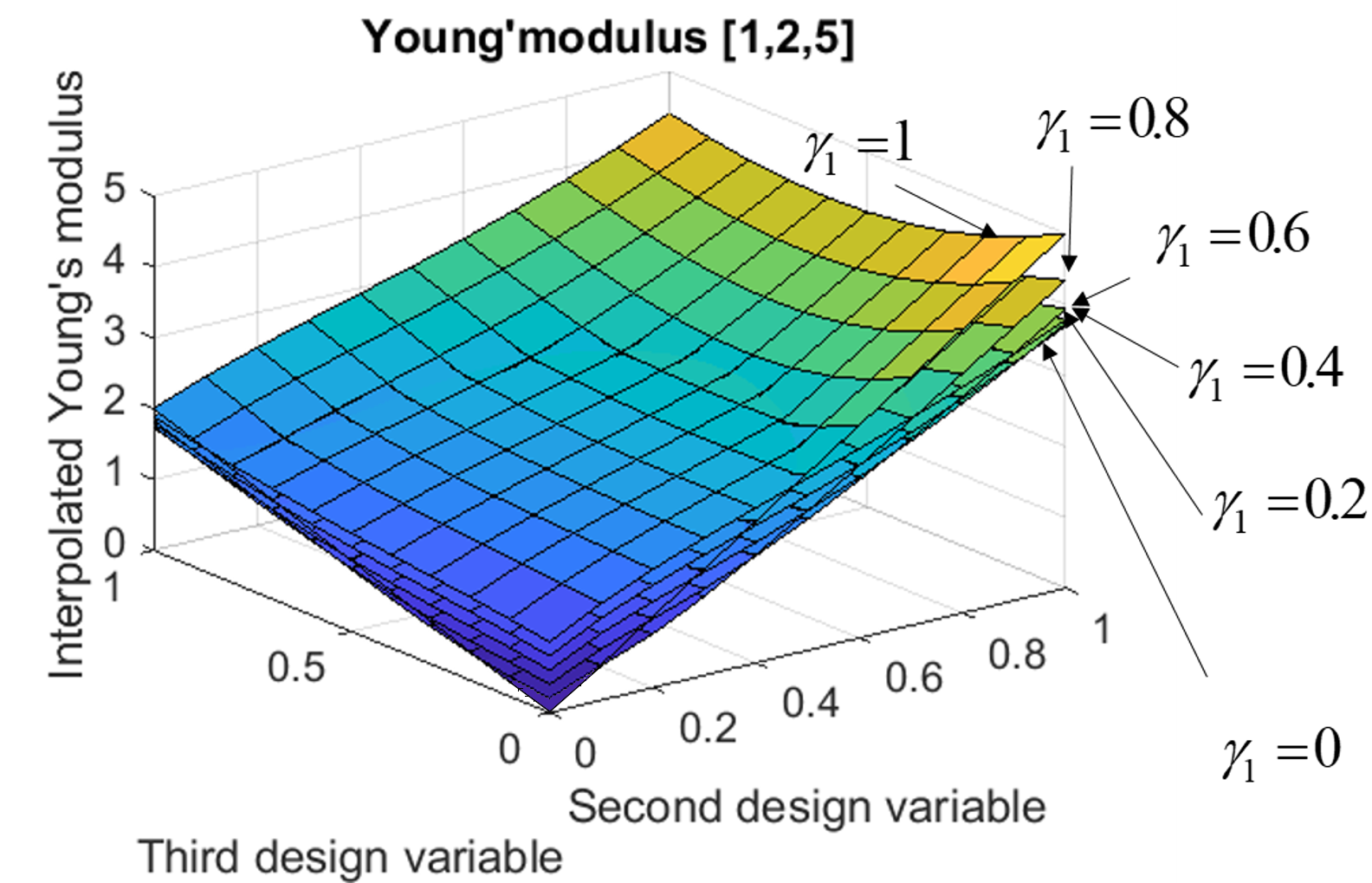}          }  \\      
  
  \subfigure[]{
     \centering
        \includegraphics[width=80mm,trim=0cm 0cm 0cm 0cm,clip]{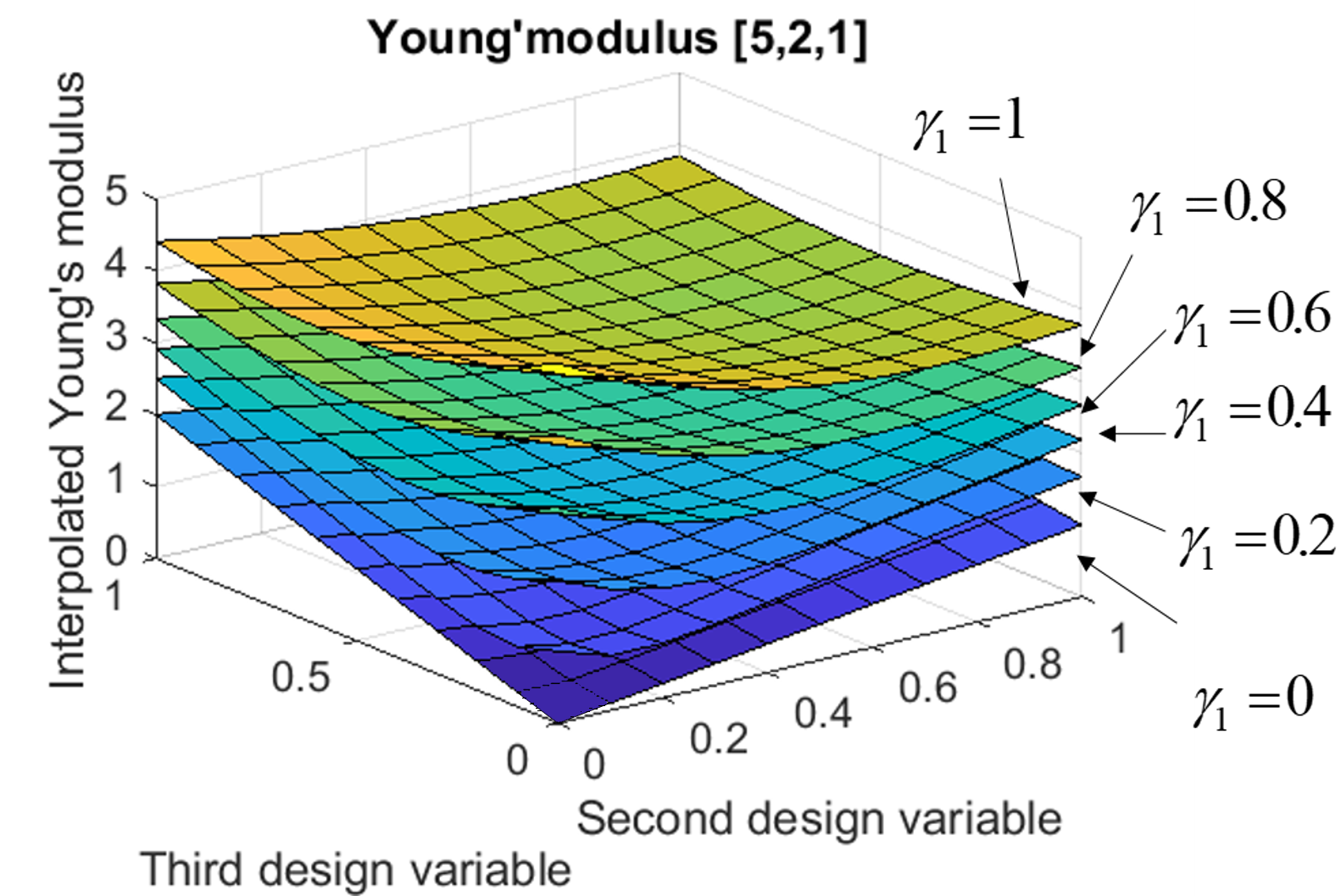}          }  \\

  \end{tabular}        
    \caption{Interpolated Young's modulus with the $p$-norm: (a) the present approach (Case 1: $n$=$p$=3, $E_1$ =5 N/m$^2$,$E_2$ =2 N/m$^2$, $E_3$ =1 N/m$^2$), (b) the present approach (Case 6: $n$=$p$=3, $E_1$ =1 N/m$^2$,$E_2$ =2 N/m$^2$, $E_3$ =5 N/m$^2$).}
    \label{fig:interpolatedYoungmodulusnew}
\end{figure}

\begin{equation}
\label{eq:stiffnessinterpolation}
\begin{gathered}
  E\left( {\gamma _1^e, \cdots ,\gamma _{NM}^e} \right) = \sum\limits_i^{NM} {{{\left( {{\phi _i}} \right)}^n}{(E_i -E_\text{void})}} + E_\text{void} \hfill \\
  {\text{where}} \hfill \\
  {\phi _i} = \frac{{{{\left\| {{\mathbf{\gamma }}_{}^e} \right\|}_{p}}}}{{{{\left\| {{\mathbf{\gamma }}_{}^e} \right\|}_{1}+\delta}}}\gamma _i^e= \frac{{{{\left\| {{\mathbf{\gamma }}_{}^e} \right\|}_{p}}}}{{\sum\limits_{} {{\mathbf{\gamma }}_{}^e} }+\delta}\gamma _i^e \hfill \\ 
  {{{\left\| {{\mathbf{\gamma }}_{}^e} \right\|}_{p}}}= (
  (\gamma _1^e)^p +(\gamma _2^e)^p  \cdots (\gamma _{NM}^e)^p)^\frac{1}{p}
\end{gathered} 
\end{equation}

where the $p$-norm and $1$-norm of the design variables of $\gamma _1^e,\gamma _2^e,..., \gamma _{NM-1}^e, \gamma _{NM}^e $, are represented by ${{\left\| {{\mathbf{\gamma }}_{}^e} \right\|}_{p}} $ and ${{\left\| {{\mathbf{\gamma }}_{}^e} \right\|}_{1}}$, respectively. A small value $\delta$ is used to avoid the undefined mathematical operation zero divided by zero. Usually, it is set as $\delta=10^{-9}$. In our formulation, the number of candidate materials is also the number of the design variables at each finite element. Therefore the total number of the design variables is the number of finite elements in the design domain times the number of candidate materials. 

For example, the following interpolation functions are set for one material, two materials and three material cases in Fig. (\ref{fig:newmappingformulation}). 

\begin{equation}
\begin{gathered}
\text{One material: } E\left( {\gamma _1^e} \right) = {\phi _1}{(\gamma _1^e)^n}(E_1^{} - E_{{\text{void}}}^{}) + E_{{\text{void}}}^{} \\
\end{gathered}
\end{equation}

\begin{equation}
\begin{gathered}
V_1=\sum\limits_{i=1}^{NE}{\gamma _1^i}
\end{gathered}
\end{equation}

\begin{equation}
\begin{gathered}
\text{Two materials: }  E\left( {\gamma _1^e,\gamma _2^e} \right) = {\phi _1}{(\gamma _1^e,\gamma _2^e)^n}(E_1^{} - E_{{\text{void}}}^{}) \hfill \\
   + {\phi _2}{(\gamma _1^e,\gamma _2^e)^n}(E_2^{} - E_{{\text{void}}}^{}) + E_{{\text{void}}}^{} \hfill \\
   \hfill \\
  {\phi _1}(\gamma _1^e,\gamma _2^e) = \frac{{{{({{(\gamma _1^e)}^p} + {{(\gamma _2^e)}^p})}^{1/p}}}}{{\gamma _1^e + \gamma _2^e} + \delta}\gamma _1^e \hfill \\
  {\phi _2}(\gamma _1^e,\gamma _2^e) = \frac{{{{({{(\gamma _1^e)}^p} + {{(\gamma _2^e)}^p})}^{1/p}}}}{{\gamma _1^e + \gamma _2^e}+\delta}\gamma _2^e \hfill \\ 
\end{gathered}
\end{equation}

\begin{equation}
\begin{gathered}
V_1=\sum\limits_{i=1}^{NE}{\gamma _1^i},V_2=\sum\limits_{i=1}^{NE}{\gamma _2^i} 
\end{gathered}
\end{equation}

\begin{equation}
\label{eq:theemateiralnew}
\begin{gathered}
\text{Three materials: }  E(\gamma _1^e,\gamma _2^e,\gamma _3^e) = \\
{\phi _1}{(\gamma _1^e,\gamma _2^e,\gamma _3^e)^n}(E_1^{} - E_{{\text{void}}}^{}) \hfill \\
   + {\phi _2}{(\gamma _1^e,\gamma _2^e,\gamma _3^e)^n}(E_2^{} - E_{{\text{void}}}^{}) \hfill \\
   + {\phi _3}{(\gamma _1^e,\gamma _2^e,\gamma _3^e)^n}(E_3^{} - E_{{\text{void}}}^{}) + E_{{\text{void}}}^{} \hfill \\
   \hfill \\
  {\phi _1}(\gamma _1^e,\gamma _2^e,\gamma _3^e) = \frac{{{{({{(\gamma _1^e)}^p} + {{(\gamma _2^e)}^p} + {{(\gamma _3^e)}^p})}^{1/p}}}}{{\gamma _1^e + \gamma _2^e + \gamma _3^e}+\delta}\gamma _1^e \hfill \\
  {\phi _2}(\gamma _1^e,\gamma _2^e,\gamma _3^e) = \frac{{{{({{(\gamma _1^e)}^p} + {{(\gamma _2^e)}^p} + {{(\gamma _3^e)}^p})}^{1/p}}}}{{\gamma _1^e + \gamma _2^e + \gamma _3^e}+\delta}\gamma _2^e \hfill \\
  {\phi _3}(\gamma _1^e,\gamma _2^e,\gamma _3^e) = \frac{{{{({{(\gamma _1^e)}^p} + {{(\gamma _2^e)}^p} + {{(\gamma _3^e)}^p})^{1/p}}}}}{{\gamma _1^e + \gamma _2^e + \gamma _3^e}+\delta}\gamma _3^e \hfill \\ 
\end{gathered} 
\end{equation}

\begin{equation}
\begin{gathered}
V_1=\sum\limits_{i=1}^{NE}{\gamma _1^i},V_2=\sum\limits_{i=1}^{NE}{\gamma _2^i}, 
V_3=\sum\limits_{i=1}^{NE}{\gamma _3^i}
\end{gathered} 
\end{equation}

The following features can be observed.  First of all, the present interpolation functions are composed with the symmetric parts with respect to the design variables multiplied with the design variable. Unlike the SIMP based interpolation function or the DMO based interpolation, they are based on the $p$-norm mapping between the space of the design variables to the space of the material properties. Therefore, regardless of the number of the involved material properties, the form of the interpolation functions are formulated systematically with respect to the design variables. With a sufficient large value for $p$, the range of this interpolation function is still between 0 and 1 that indicates nonphysical material properties not allowed, which follows the requirement of the Hashin-Shtrikman bounds for multiphase composites or multi-material (\cite{HASHIN1963127}). With a larger value for $p$, a higher penalization can be achievable, and a clearer optimized structure with multiple material can be obtained. In practices, we set $p=6$ to balance the clear optimized structure and the robustness of the optimizer.

%Figure. \ref{fig:twothreecomparison} compares the functional spaces of the present interpolation function. 

Secondly, the interpolated Young's modulus simply becomes that of the $e$-th material with zeros for all the design variables and with one for the $e$-th design variable. The present interpolation function in  Fig. \ref{fig:interpolatedYoungmodulusnew} also relies on the characteristics of an optimizer to make the design variables converged to the limits, i.e., 0 or 1. Similar to the SIMP based interpolation function with an increasing order of the Young’s modular for each material Fig.\ref{fig:interpolatedYoungmodulus}(b),and any finite element with intermediate values can be penalized since the constitutive properties with intermediate variables becomes uneconomic. But for the SIMP based interpolation function with a decreasing order of the Young’s modular for each material Fig.\ref{fig:interpolatedYoungmodulus}(a)and the DMO based method with both of these cases Fig.\ref{fig:interpolatedYoungmodulus}(c) and (d), are difficult to converge to clear result.  

In addition, the setting of the initial values of the design variables becomes straight forward.  For an example, we can easily set the design variable to 0.5 to control the initial amount of each material to be 0.5/3 with controllable and equivalent initialization for each material according to Equation.(\ref{eq:theemateiralnew}) , but it is difficult to control with both the SIMP and DMO based interpolation function. Finally, it is also easy to implement and extend the present mapping based interpolation for topology optimization of much more materials.

\subsection{Filtering and Projection}
\subsubsection{Filtering}
In order to avoid the formation of checker-board patter for the topology optimization problem, two types of filtering methods are used, naming the simple sensitivity filter and the density filter based on the Helmholtz type partial differential equation (PDE).

\textbf{Sensitivity filter:} A common approach is the application of a filter to the sensitivities of each element based on their neighbourhood (\cite{Andreassen2011}). Hence, The sensitivities of the objective and the constraint with respect to the ${\mathbf{\gamma}_{i}^{e}}$ design variable can be modified as follows:

\begin{equation}
\widehat{\frac{\partial c}{\partial  {\mathbf{\gamma}_{i}^{e}}}} =  \frac{\sum_{e\in N_{e}}\frac{\partial c}{\partial  {\mathbf{\gamma}_{i}^{e}}}}{\sum_{e\in N_{e}} 1}
\end{equation}

where $N_{e}$ is the set of neighbourhood of each element $e$ with a given influence radius $R$.

\textbf{Density filter:} The density filter based on the Helmholtz type partial differential equation (PDE) with homogeneous Neumann boundary conditions (\cite{Andreassen2011}) is also introduced for each material, which can be expressed as:

\begin{equation}
-R_{min}^{2}\nabla ^{2}\widetilde{\gamma_{i}}+\widetilde{\gamma_{i}}=\gamma_{i}
\end{equation}

\begin{equation}
\frac{\partial \widetilde{\gamma_{i}}}{\partial \textbf{n}}=0
\end{equation}

where $\widetilde{\gamma_{i}}$ is the filtered field of the design variable for each material. The parameter $R_{min}$ plays a similar roles as $R$ for the simple sensitivity filter based on the average weight of the radius of the neighbourhood. An approximate relation between them is $R_{min}=2\sqrt{3}R$.

\subsubsection{Projection}
The filtered field of the design variable for each material $\widetilde{\gamma_{i}}$ contains grey elements, the projection method is used to ensure 0-1 solution, which can be written as:

\begin{equation}
\overline{\gamma_{i}}=\frac{tanh(\frac{\beta}{2})+tanh(\beta(\widetilde{\gamma_{i}}-\frac{1}{2}))}{2tanh(\frac{\beta}{2})}
\end{equation}

The parameter $\beta$ controls the sharpness of the projection function. We start with a small value of $\beta$ and double its value after a certain number of iterations.
%%%%%%%%%%%%%%%%%%%%%%%%
%\input{section3}
\section{Optimization results}
\label{sec3}

To prove the concept of topology optimization for multiple materials with the mapping based interpolation, this section solves two dimensional compliance minimization problems.  Finite element implementation with the mapping based interpolation function, the sensitivity analysis and the optimization were implemented in the framework of the Matlab. To solve the optimization problem, the method of moving asymptotes (MMA) algorithm \citep{svanberg87} is implemented. 

\subsection{Example 1: Cantilever beam problem}

\begin{figure}
\centering
\includegraphics[scale=0.5]{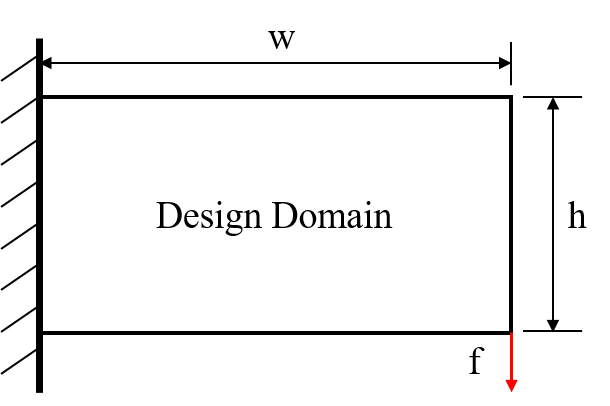} 
    \caption{Design domain and boundary condition for the cantilever beam example}
    \label{fig:cantilverExp}
\end{figure}

For the first example, the cantilever beam problem in Fig.~\ref{fig:cantilverExp} is solved with the present mapping based interpolation function. The design domain defined as a rectangle area of unit thickness with width $w=2$ and height $h=1$, and is fixed at the left side. A concentrated load $f=0.1$ is applied at the lower right vertex of the rectangle. The design domain is discretized into 200 by 100 equally-sized square four-node element $r_e=0.01$ with Young’s modulus $E=1$ and Poisson’s ratio $\mu=0.3$. A simple sensitivity filter based on the average weight of the radius $R=1.5r_e$ of the neighbourhood (\cite{Andreassen2011}) is used to solve the check-board problem.

\begin{figure}
\centering
    \subfigure[]{
     \centering
        \includegraphics[scale=0.13]{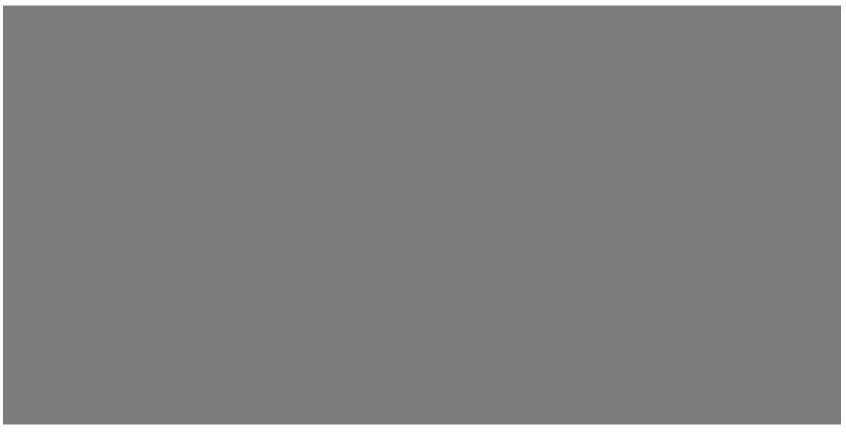}
          \label{(a)}} 
    \subfigure[]{
    \centering
        \includegraphics[scale=0.13]{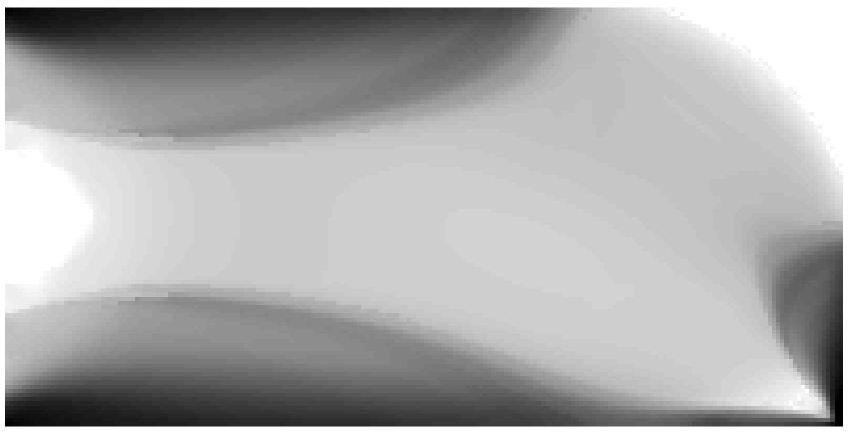}
          \label{(b)}} 
     \subfigure[]{
    \centering
\includegraphics[scale=0.13]{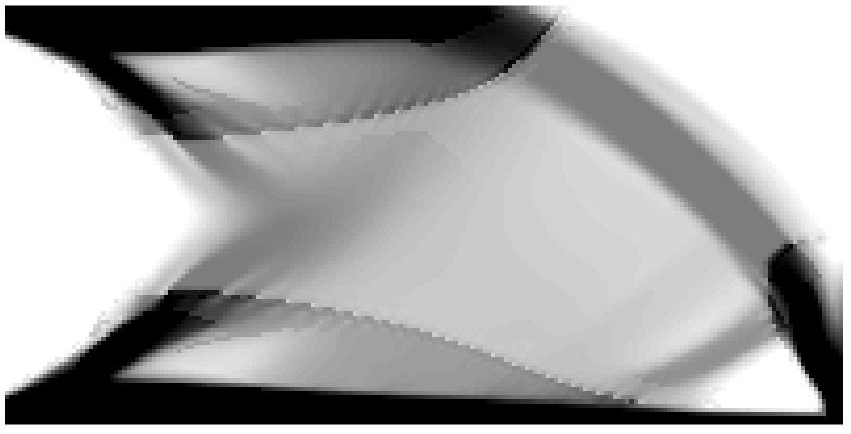}
          \label{(c)}} 
     \subfigure[]{
    \centering
\includegraphics[scale=0.13]{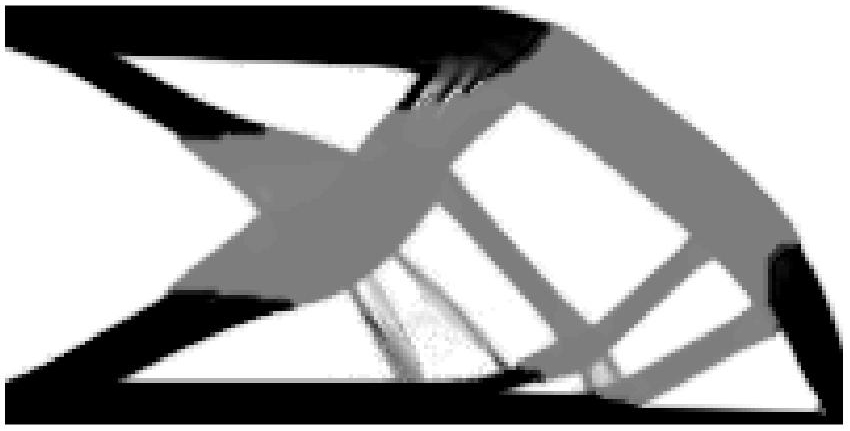}
          \label{(d)}}  
              \subfigure[]{
     \centering
        \includegraphics[scale=0.13]{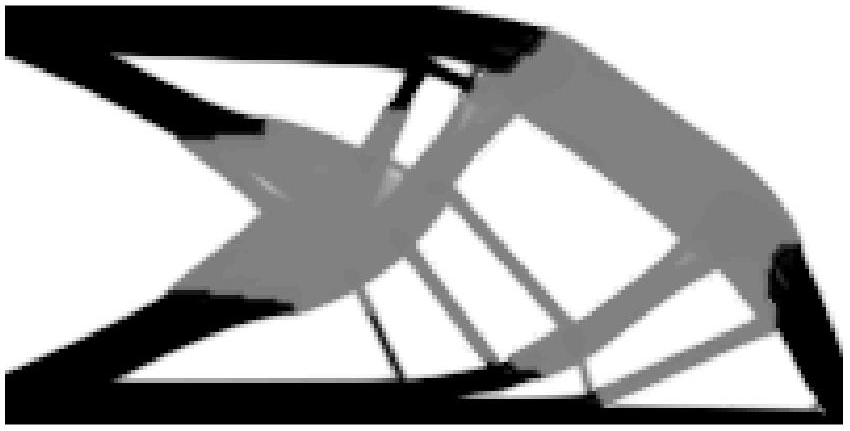}
          \label{(e)}} 
    \subfigure[]{
    \centering
        \includegraphics[scale=0.13]{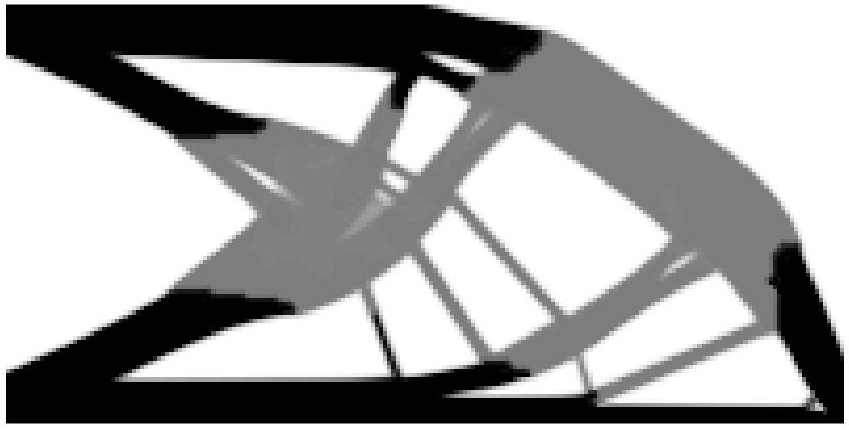}
          \label{(f)}} 
    \caption{The iteration details of Cantilever with 2 materials for the mapping based method. (a) 1st iterations, (b) 10th iterations,(c) 20th iterations, (d) 40th iterations, (e) 80th iterations, (f) 200th iterations}
\label{fig:itersSnopts}
\end{figure}

\begin{figure}
\centering
\includegraphics[scale=0.4]{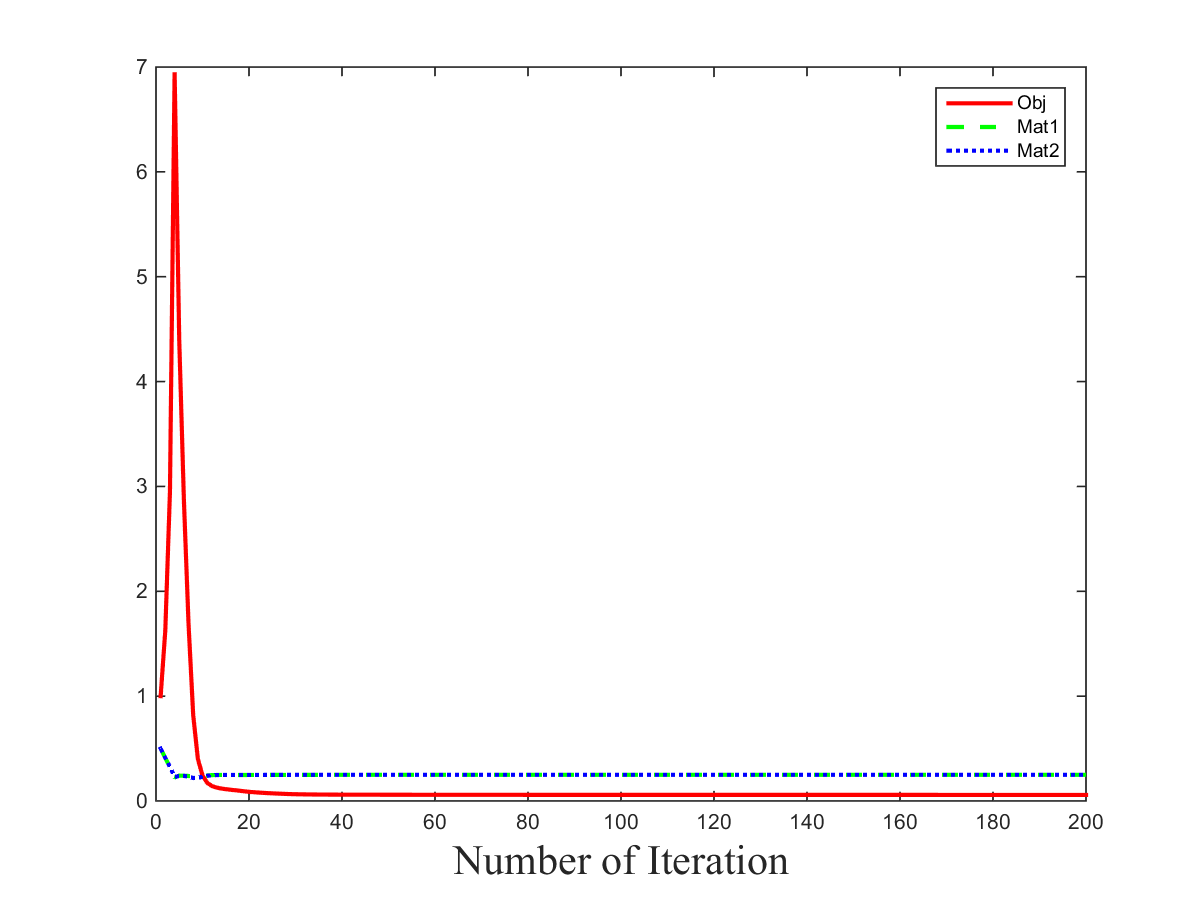} 
    \caption{The converge history of topology optimization of Cantilever with 2 materials}
    \label{fig:conv_iter}
\end{figure}

Figure \ref{fig:itersSnopts} shows the snapshots for the convergence history of the mapping based topology optimization of multi-materials at different iterations during optimization. The Young's modular for the stiff materials and compliant materials are set to be 5 and 1, respectively. The density field is initialized equality and uniformly with 0.5 for each material over the entire design domain shown in Fig.\ref{fig:itersSnopts}(a), which is impossible for convectional SIMP and DMO interpolation function. The results for 10th, 20th, 40th, 80th and 200th iterations are shown in Fig.\ref{fig:itersSnopts}(b), (c), (d), (e), (f), respectively. It can be obviously found that the proposed method almost converged to the final result only with 40 iterations, as shown in Fig.\ref{fig:itersSnopts}(d). With the iteration number is increasing, only a few change of the part that is near the interfaces between materials occur, and the boundary of each material become much more clear without any overlap. The convergence history of the optimization shown in Fig.\ref{fig:conv_iter} also proves the fast converging point, as is shown in the plot of the object function with the red solid line. The plot of the volume convergence history of the compliant material and the stiff material are shown with green dash dot line and blue dot line, respectively. It can also obviously show that both materials are initialized with the same value of 0.5. The volume of each material decreases quickly while the object function increasing at the first few iterations. Then, both the stiff and compliant material would be constraint with 0.25 as expected within 20 iterations while the object function decreasing.

\begin{figure}
\centering
    \subfigure[]{
     \centering
        \includegraphics[scale=0.105]{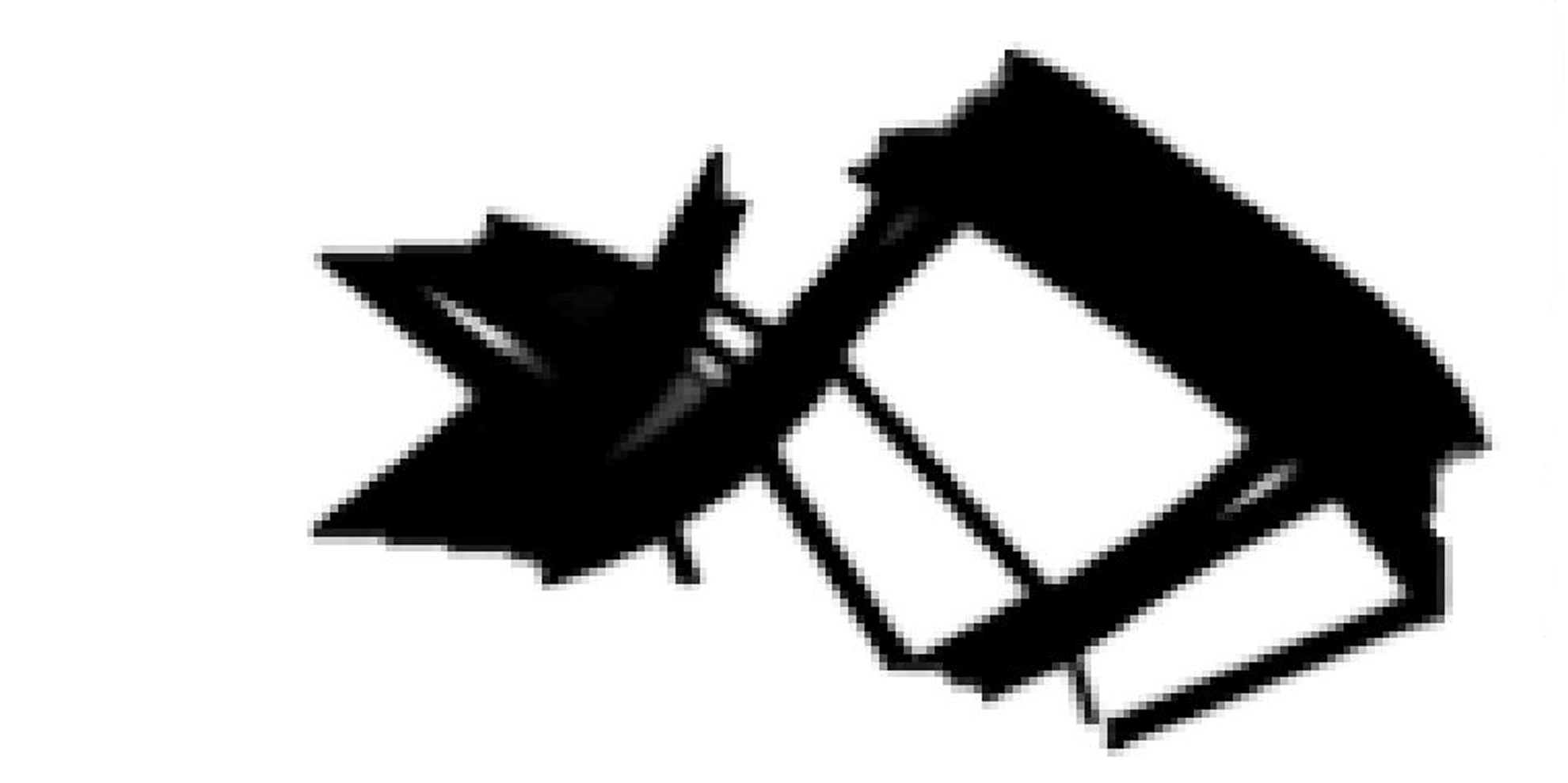}
          \label{(a)}} 
    \subfigure[]{
    \centering
        \includegraphics[scale=0.105]{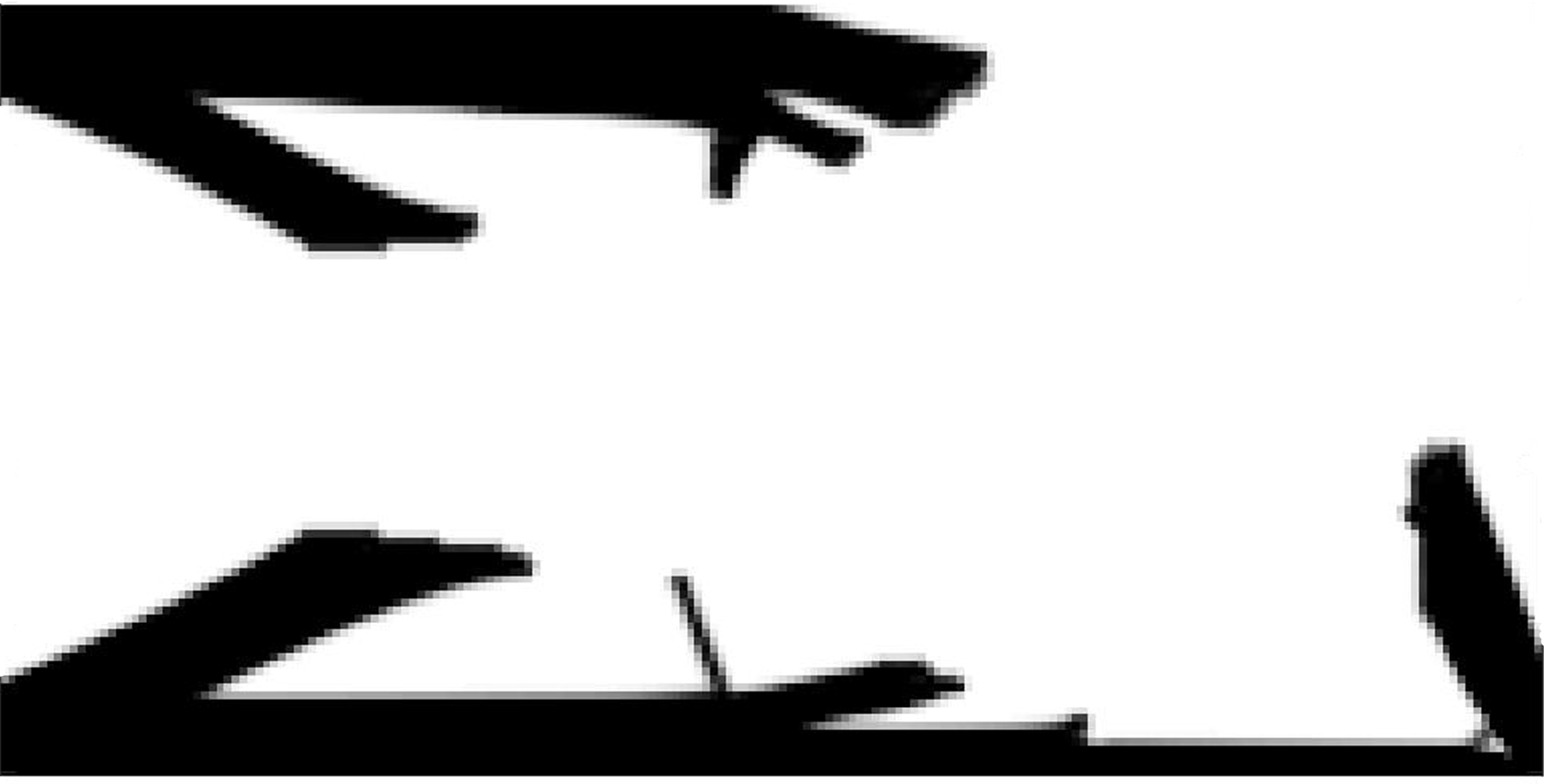}
          \label{(b)}} 
     \subfigure[]{
    \centering
    \includegraphics[scale=0.25]{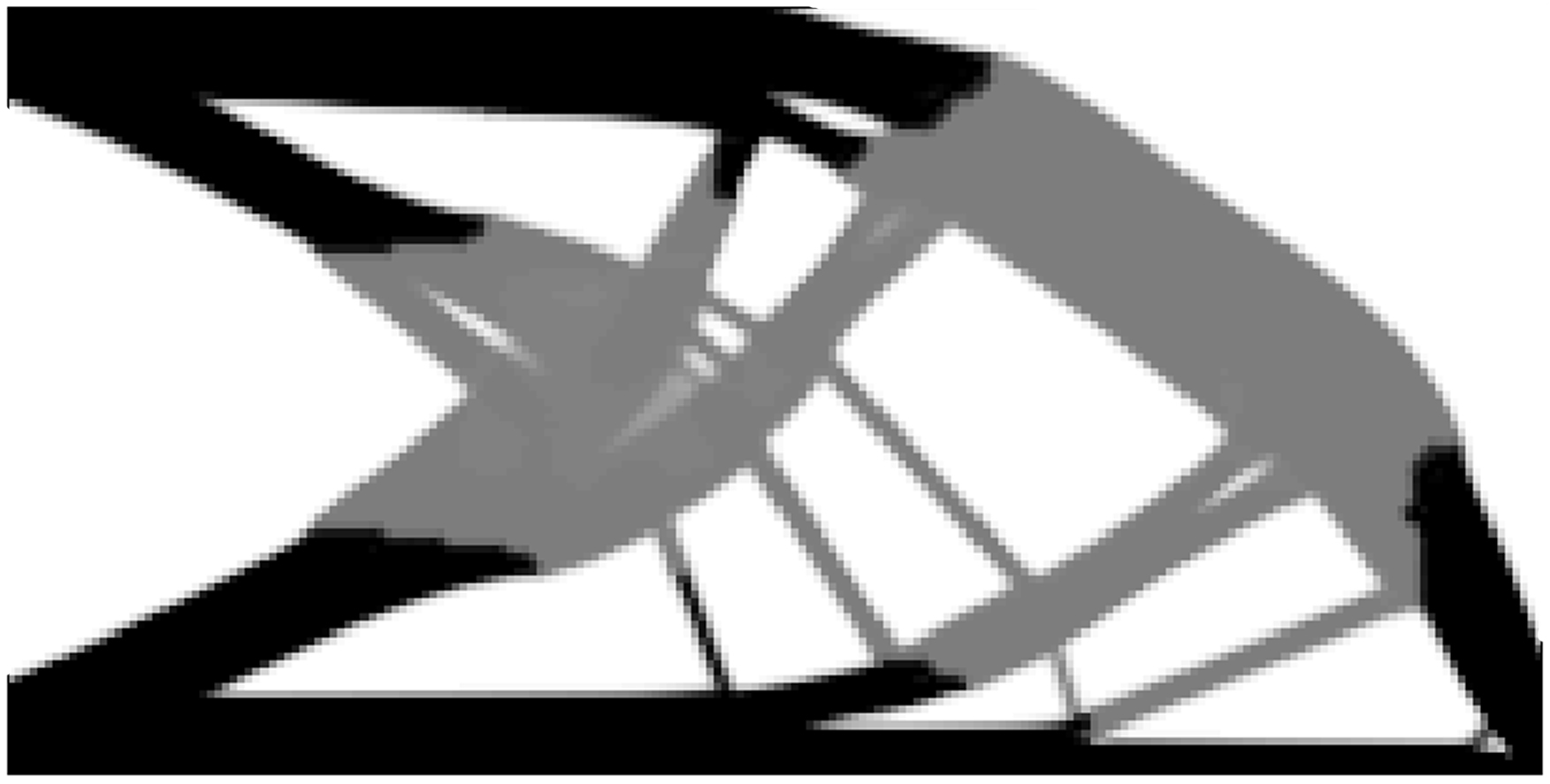}
          \label{(c)}} 
    \caption{Topology optimization of Cantilever with 2 materials, $V_1=V_2=0.25$, compliance = 0.451 (a) material 1, $E=1$ (b) material 2, $E=5$ (c) total structure( the black is the stiff material and the grey is the compliant material) }
    \label{fig:multimaterialTO2}
\end{figure}

The details of the optimized structures are shown in Fig.~\ref{fig:multimaterialTO2}. The compliant and stiff materials are shown Fig.~\ref{fig:multimaterialTO2}(a) and (b), respectively. The whole structure is shown in Fig.~\ref{fig:multimaterialTO2}(c), different colors represent different materials, the black is the stiff material and the grey is the compliant material. The compliance of the optimized structure is 0.451. One interesting thing can be found that the boundaries between each material are clear. It is because that the presented paper can get only 0, 1 result for each component without any overlap with others. However, most of conventional methods such the extended SIMP based method will produce smooth transit between each material and even with overlap near the boundary, which would decrease the structure stiffness when put into real manufacturing. This is one of the big advantages of the proposed method over conventional methods, which will be discussed in details in the comparison section. 

\begin{figure}
\centering
\includegraphics[scale=0.5]{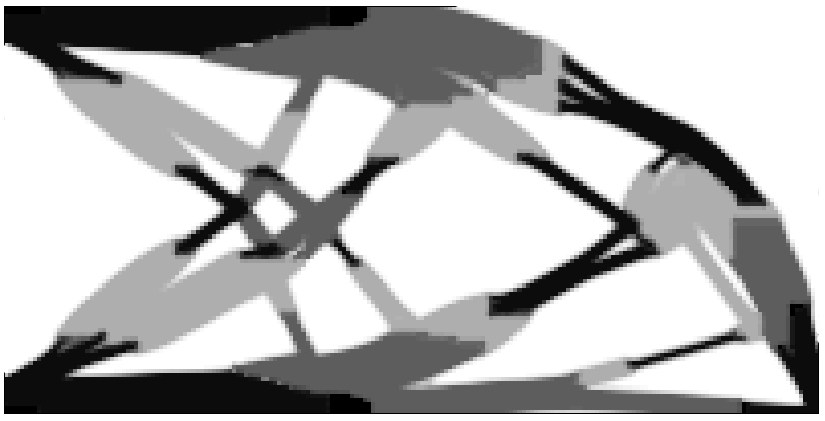} 
    \caption{Topology optimization of Cantilever with 3 materials, different colors represent different materials, $E_1=1, E_2=2, E_3=5$, $V_1=V_2=V_3=\frac{0.5}{3}$, compliance = 0.252}
    \label{fig:multimaterialTO3}
\end{figure}

\begin{figure}
\centering
\includegraphics[scale=0.5]{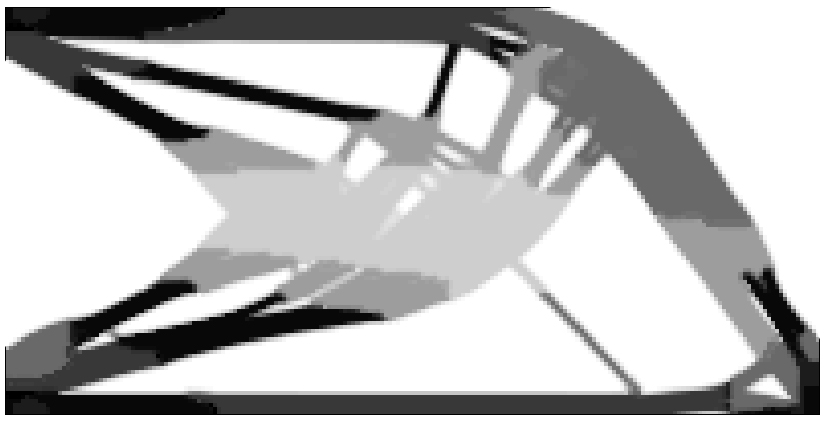} 
    \caption{Topology optimization of Cantilever with 5 materials, different colors represent different materials, $E_1=1, E_2=2, E_3=3, E_4=4, E_5=5$, $V_1=V_2=V_3=V_4=V_5=0.1$, compliance = 0.344}
    \label{fig:multimaterialTO5}
\end{figure}

Figure.~\ref{fig:multimaterialTO3}, Figure.~\ref{fig:multimaterialTO5} also show the optima layouts with three and five materials with the volume constraint of 0.5/3 and 0.1 for each material, respectively. It can be obviously found out that the stiff materials appear at the boundary condition, the outer domain and the domain at the loading point for each scenery. The Young's modular are set to 1, 2, 5 for three materials, and 1, 2, 3, 4, 5 for five materials, respectively. The material property of the void region is set to a small number $1e^{-9}$. The initial guess is set to 0.5 for each material. With the presented mapping based interpolation function, the values of the interpolation functions with the uniform initial values are able to be set as the same. The compliance for 3 and 5 material is 0.252 and 0.344, respectively. It can be found that the compliance vary with the number of materials used for optimization even with the same up and low bound of Young's modular for the materials. It is proved that the structure stiffness to weight ratio can be improved by using multi-materials. According to the simple formulation of the proposed method, is also easy to achieve the topology optimization of a large number of materials as the computation source is enough.  

\begin{figure}
\centering
  \begin{tabular}{cc}
   \subfigure[]{
     \centering
        \includegraphics[width=38mm]{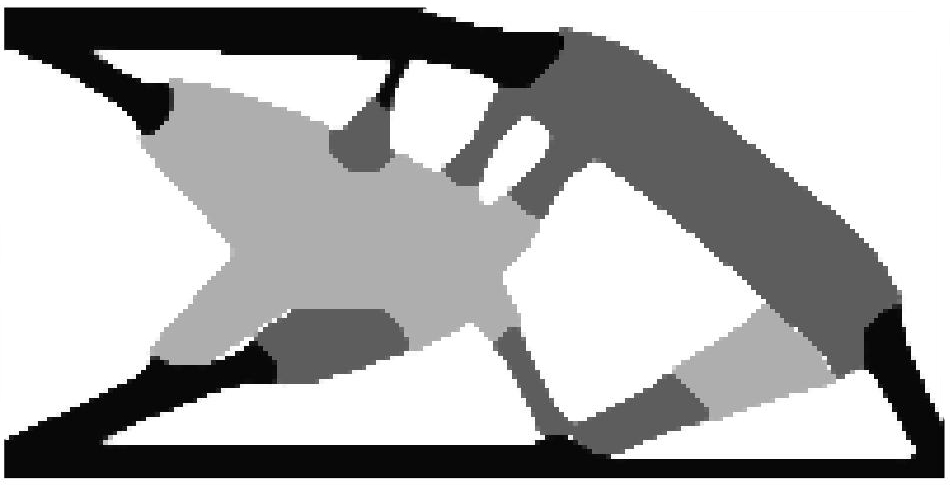}}
    \subfigure[]{
     \centering
        \includegraphics[width=38mm]{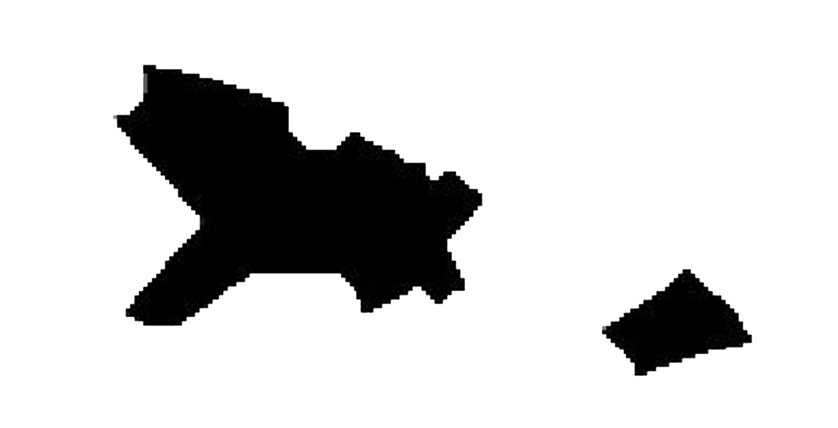}          }\\        
    \subfigure[]{
        \includegraphics[width=38mm]{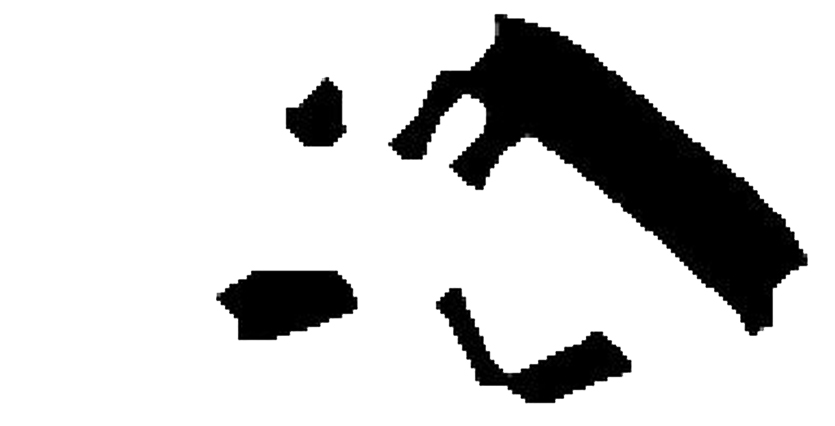}          }
  \subfigure[]{
     \centering
        \includegraphics[width=38mm]{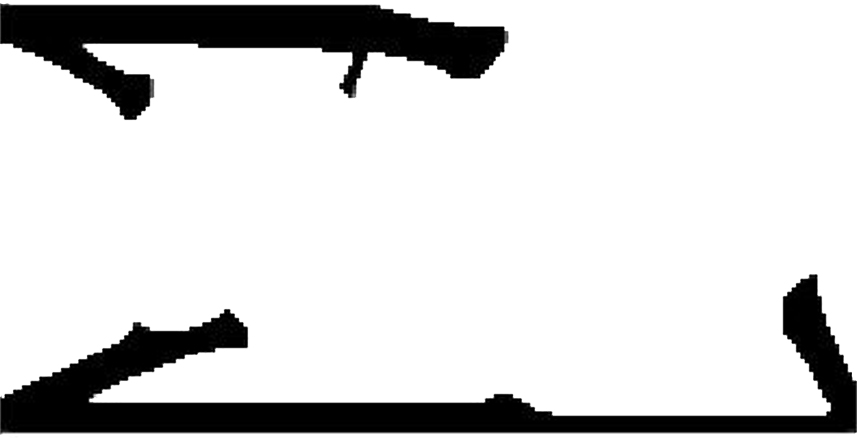}          }\\       
  \end{tabular}        
    \caption{Topology optimization of Cantilever with 3 materials with $p=3$, $V_1=V_2=V_3=\frac{0.5}{3}$, compliance = 0.229 (a) total structure, different colors represent different materials (b) material 1, $E=1$ (c) material 2, $E=2$ (d) material 3, $E=5$ }
    \label{fig:TOP_cat_m3_p3}
\end{figure}

\begin{figure}
\centering
  \begin{tabular}{cc}
   \subfigure[]{
     \centering
        \includegraphics[width=38mm]{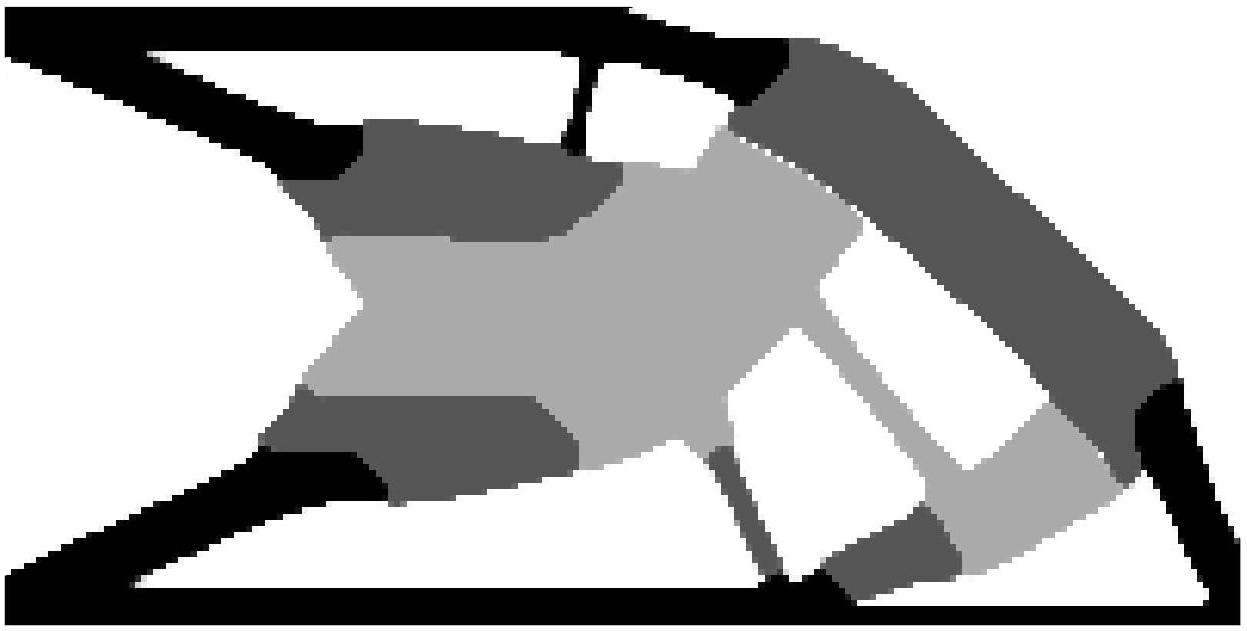}}
    \subfigure[]{
     \centering
        \includegraphics[width=38mm]{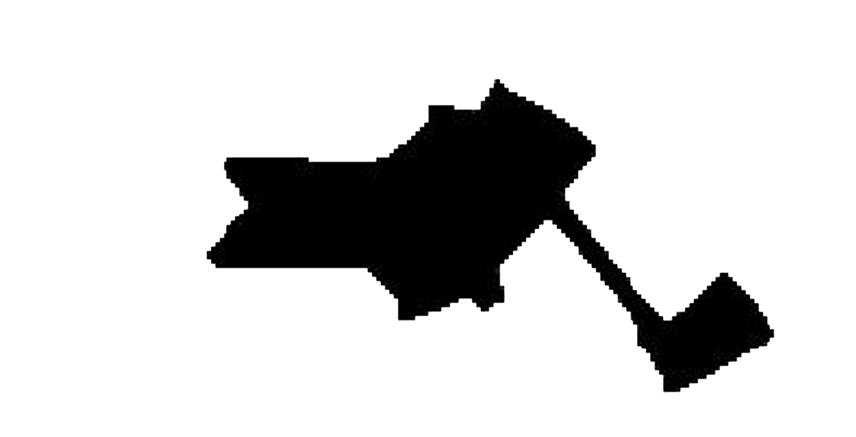}          }\\        
    \subfigure[]{
        \includegraphics[width=38mm]{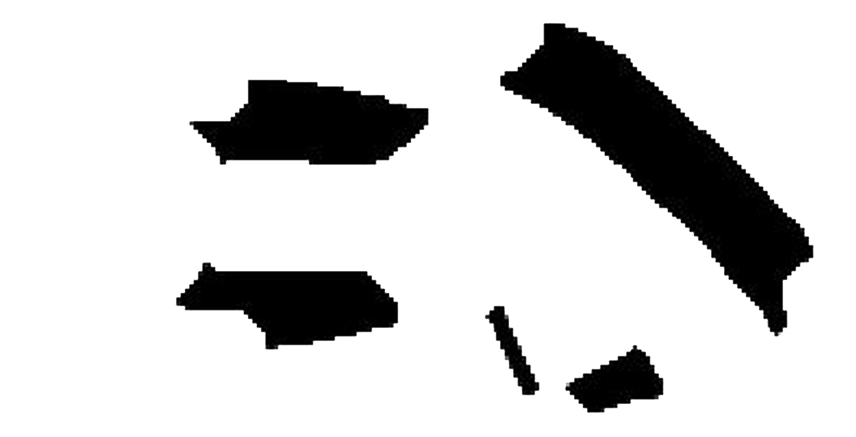}          }
  \subfigure[]{
     \centering
        \includegraphics[width=38mm]{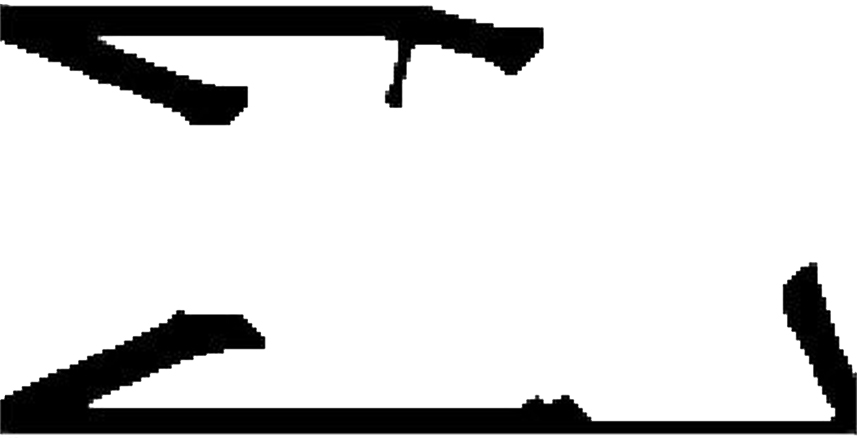}          }\\       
  \end{tabular}        
    \caption{Topology optimization of Cantilever with 3 materials with $p=6$, $V_1=V_2=V_3=\frac{0.5}{3}$, compliance = 0.228 (a) total structure, different colors represent different materials (b) material 1, $E=1$ (c) material 2, $E=2$ (d) material 3, $E=5$ } 
    \label{fig:TOP_cat_m3_p6}
\end{figure}

\begin{figure}
\centering
  \begin{tabular}{cc}
   \subfigure[]{
     \centering
        \includegraphics[width=38mm]{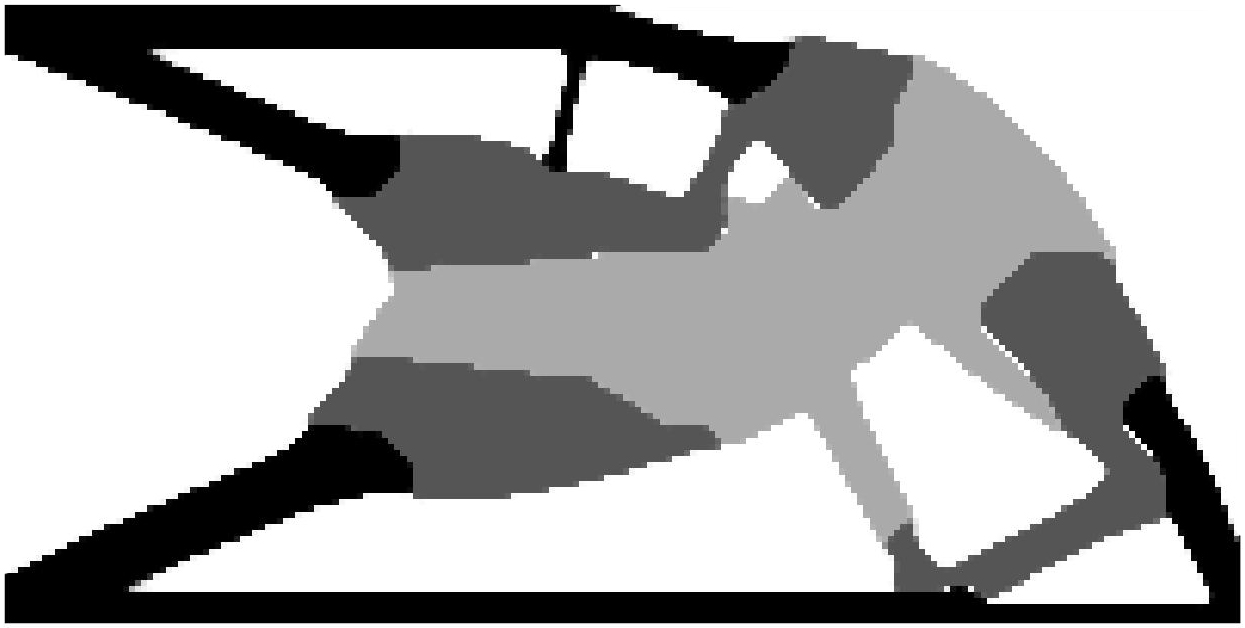}}
    \subfigure[]{
     \centering
        \includegraphics[width=38mm]{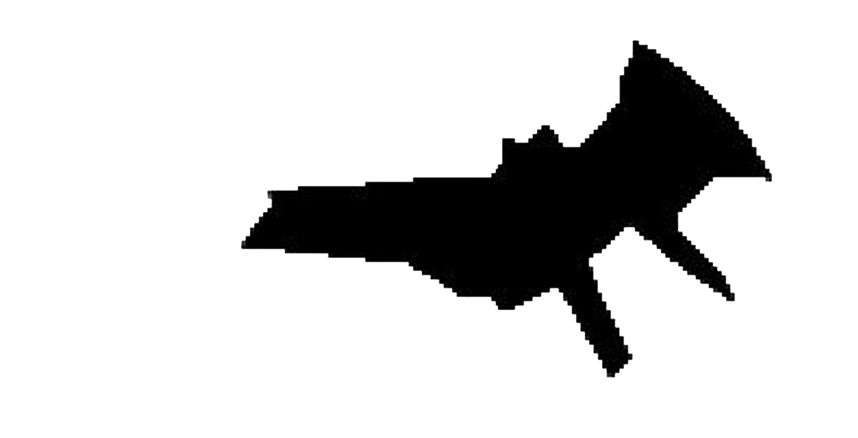}          }\\        
    \subfigure[]{
        \includegraphics[width=38mm]{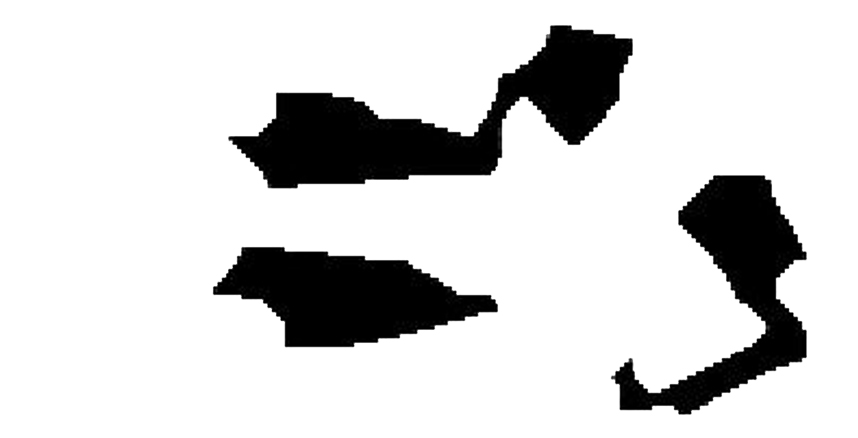}          }
  \subfigure[]{
     \centering
        \includegraphics[width=38mm]{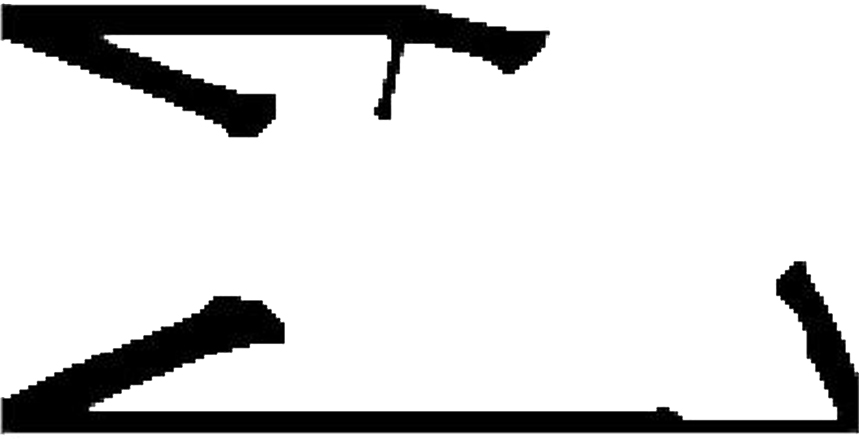}          }\\       
  \end{tabular}        
    \caption{Topology optimization of Cantilever with 3 materials with $p=16$, $V_1=V_2=V_3=\frac{0.5}{3}$, compliance = 0.234 (a) total structure, different colors represent different materials (b) material 1, $E=1$ (c) material 2, $E=2$ (d) material 3, $E=5$ }
    \label{fig:TOP_cat_m3_p16}
\end{figure}

We also implemented the proposed method by using the Helmholtz type PDE filter method (\cite{Andreassen2011})  and the approximate differentiable projection method (\cite{Wu7829422}) to explore the influence of the value of $p$ with the p-norm formulation to the optimized structure. The parameter $R_{min} = \frac{4r}{2\sqrt{3}}$ is set for the PDE filter method. To improve the convergence behavior, the common technique which is known as parameter continuation is used, we start with $\beta=2$ and double its value after 50 iterations. The optimized structures of the proposed method with $p=3$, $p=6$, $p=16$ are shown in Fig.\ref{fig:TOP_cat_m3_p3}, Fig.\ref{fig:TOP_cat_m3_p6}, Fig.\ref{fig:TOP_cat_m3_p16}, respectively. The proposed method converges to clear 0,1 result without grey element of each material for all the cases. It also can be found that the optimizer fill strong material near the periphery of the structure where stiffness is needed, and obtain a comparable stiff structure. We recommend to use $p=6$ for the proposed method to balance the stiff structure and the robustness of the optimizer. 

\subsection{Example 2: MBB  problem}

\begin{figure}
\centering
\includegraphics[scale=0.5]{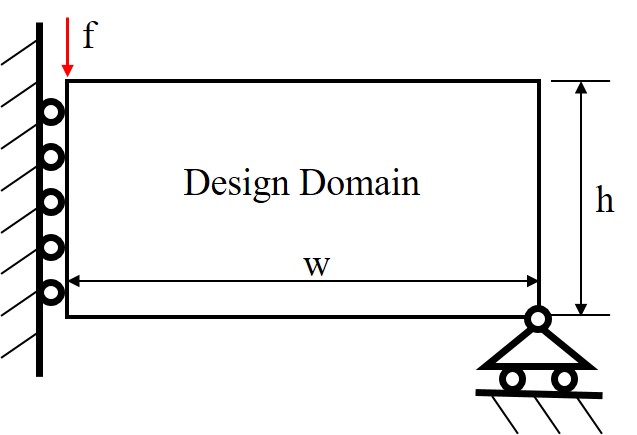} 
    \caption{Design domain and boundary condition for the half MBB example}
    \label{fig:MBBExp}
\end{figure}

The half MBB problem shown in Fig. \ref{fig:MBBExp} is considered for the second problem. The design domain is defined as a rectangle area of unit thickness with width $w=2$ and height $h=1$, and it is simply-supported at the bottom corners. A concentrated load $f=1$ is applied at the  left upper side of the rectangle and the symmetric constraint is applied at the left side. The design domain is discretized into 200 by 100 equally-sized square four-node element $r_e=0.01$. The Poisson’s ratio, the volume constraint for each material, and the filtering method are set as the same with the first example.

\begin{figure}
\centering
\includegraphics[scale=0.5]{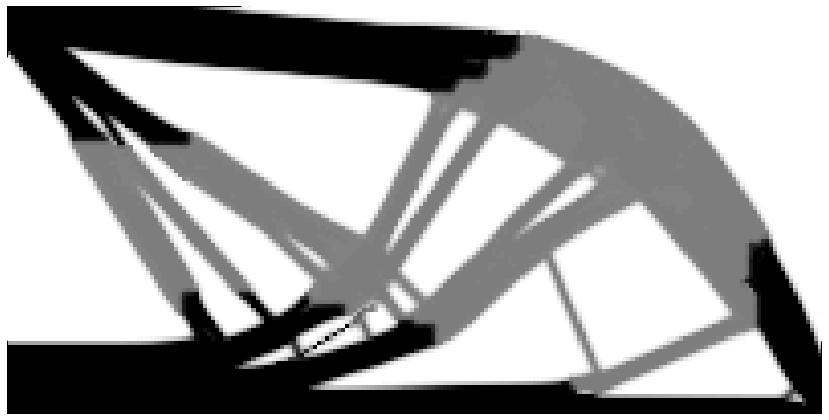} 
    \caption{Topology optimization of MBB with 2 materials, compliance = 0.507(different colors represent different materials)}
    \label{fig:multimaterialTOmbb2}
\end{figure}

\begin{figure}
\centering
\includegraphics[scale=0.5]{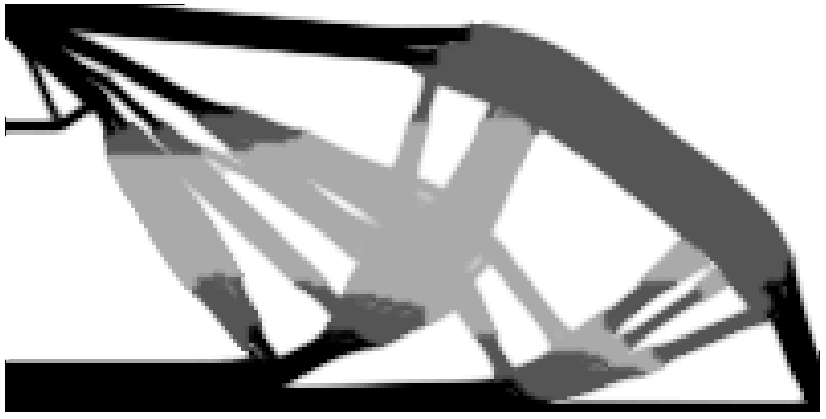} 
    \caption{Topology optimization of MBB with 3 materials, compliance = 0.279(different colors represent different materials)}
    \label{fig:multimaterialTOmbb3}
\end{figure}

\begin{figure}
\centering
\includegraphics[scale=0.5]{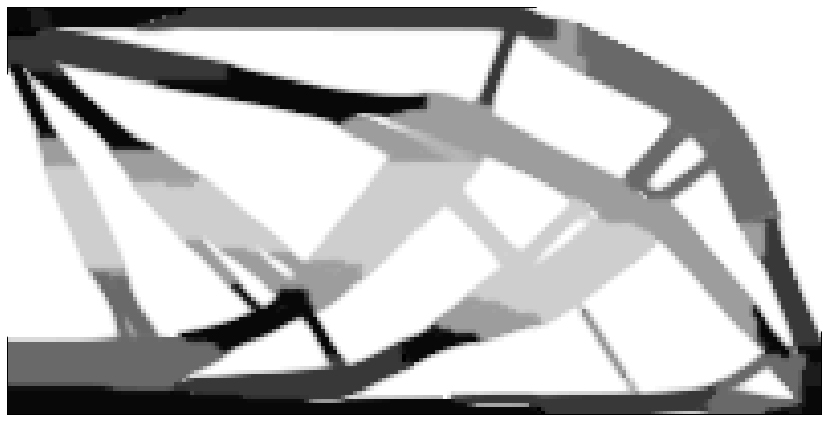} 
    \caption{Topology optimization of MBB with 5 materials, compliance = 0.333(different colors represent different materials)}
    \label{fig:multimaterialTOmbb5}
\end{figure}

The results for MBB with 2, 3, and 5 materials are shown in Fig.~\ref{fig:multimaterialTOmbb2}, Fig.~\ref{fig:multimaterialTOmbb3}, Fig.~\ref{fig:multimaterialTOmbb5}, respectively. Similarly, It can be found that the optimizer fills strong material near the periphery of the structure where stiffness is needed. The compliance for 2, 3 and 5 material are 0.507, 0.279 and 0.333, respectively. It also shows that we need to balance the number of the material to achieve a stiffer structure with the same total volume constraint. The results also prove the clear 0, 1 value for each material without overlap or smooth constraint, which would make the simultaneous physical performance for the optimization and manufactured structure.

\begin{figure}
\centering
  \begin{tabular}{cc}
   \subfigure[]{
     \centering
        \includegraphics[width=38mm]{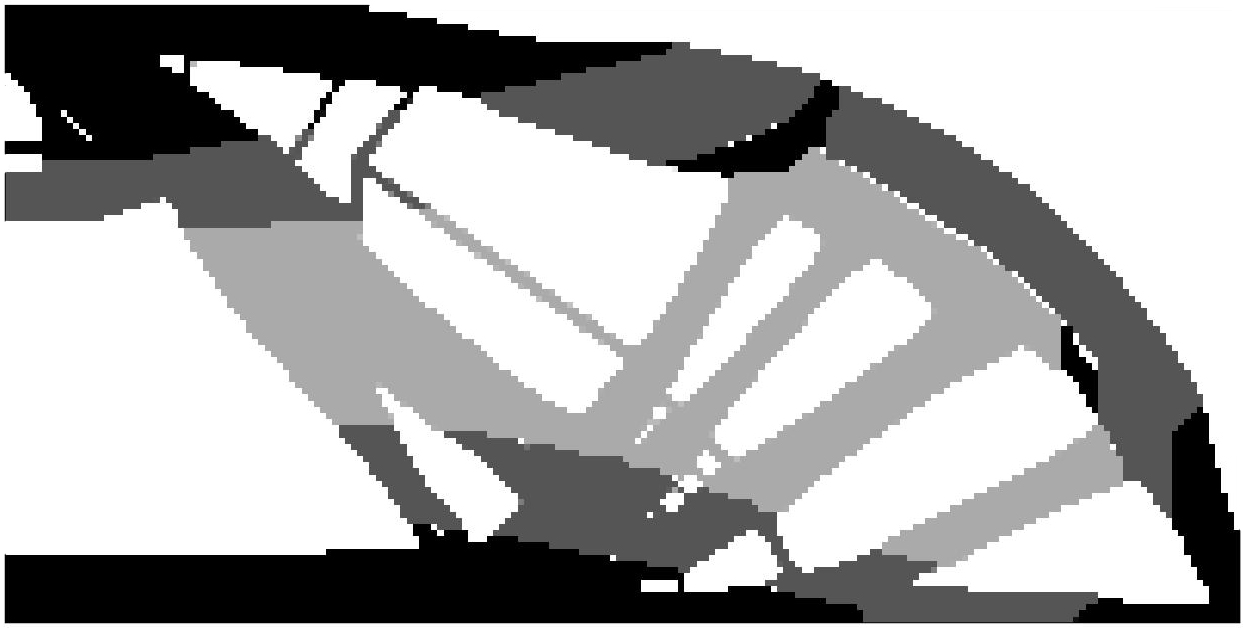}}
    \subfigure[]{
     \centering
        \includegraphics[width=38mm]{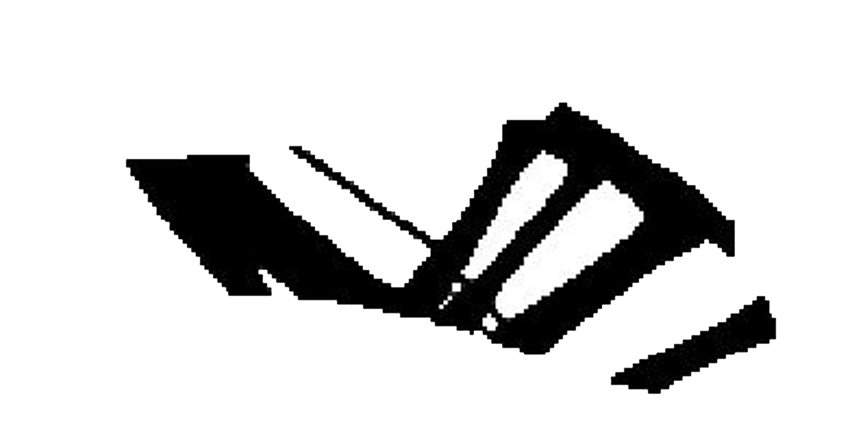}          }\\        
    \subfigure[]{
        \includegraphics[width=38mm]{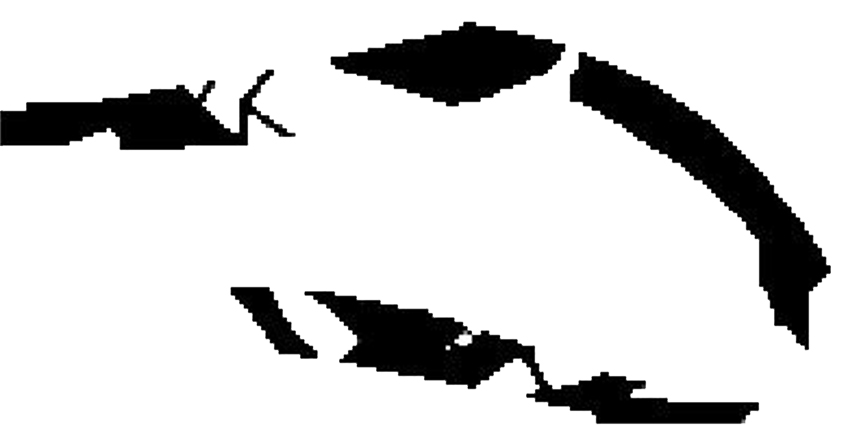}          }
  \subfigure[]{
     \centering
        \includegraphics[width=38mm]{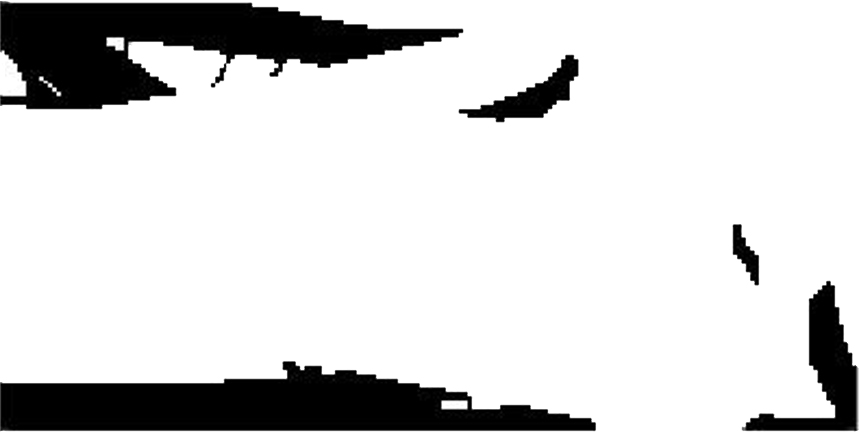}          }\\       
  \end{tabular}        
    \caption{Topology optimization of MBB with 3 materials with $R_{min}=\frac{2r_e}{2\sqrt{3}}$, $V_1=V_2=V_3=\frac{0.5}{3}$, compliance = 0.269 (a) total structure, different colors represent different materials (b) material 1, $E=1$ (c) material 2, $E=2$ (d) material 3, $E=5$ } 
    \label{fig:TOP_mbb_m3_r2}
\end{figure}

\begin{figure}
\centering
  \begin{tabular}{cc}
   \subfigure[]{
     \centering
        \includegraphics[width=38mm]{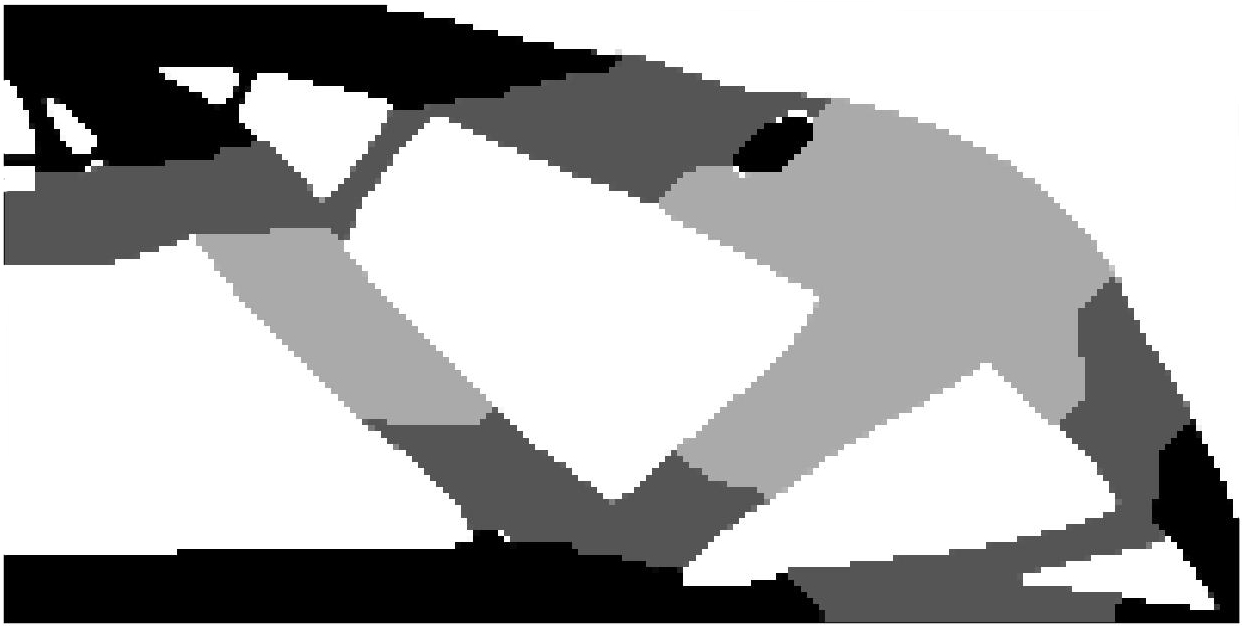}}
    \subfigure[]{
     \centering
        \includegraphics[width=38mm]{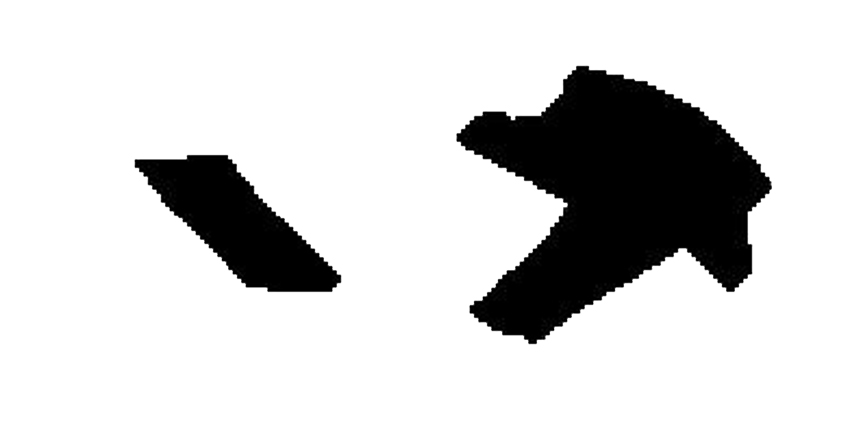}          }\\        
    \subfigure[]{
        \includegraphics[width=38mm]{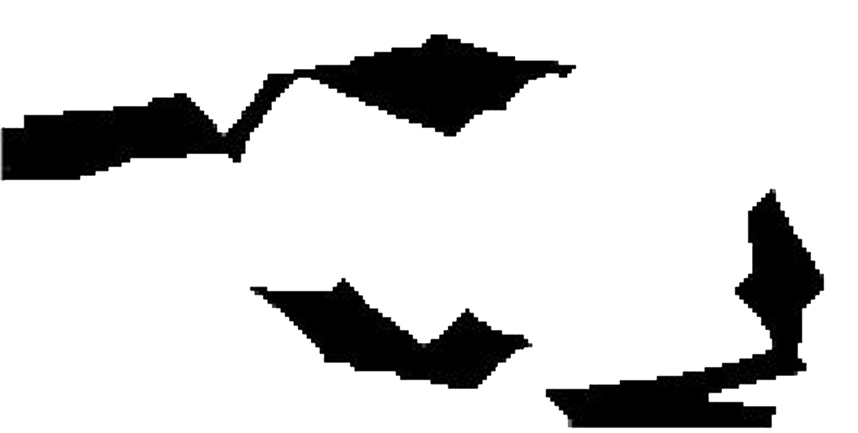}          }
  \subfigure[]{
     \centering
        \includegraphics[width=38mm]{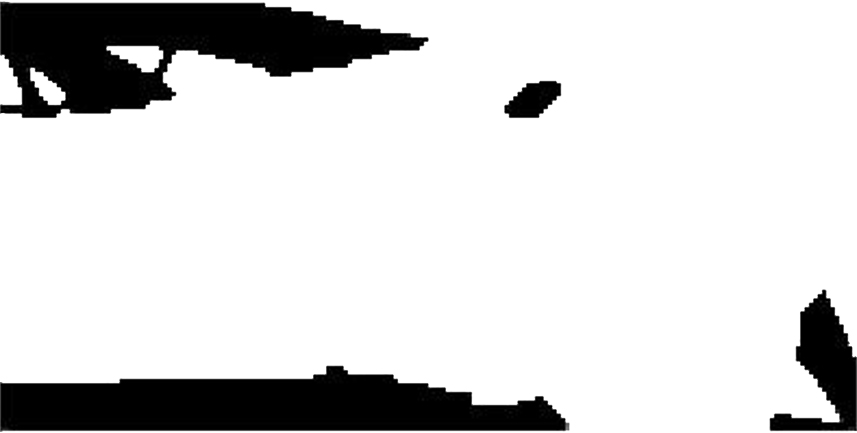}          }\\        
  \end{tabular}        
    \caption{Topology optimization of MBB with 3 materials with $R_{min}=\frac{4r_e}{2\sqrt{3}}$, $V_1=V_2=V_3=\frac{0.5}{3}$, compliance = 0.280 (a) total structure, different colors represent different materials (b) material 1, $E=1$ (c) material 2, $E=2$ (d) material 3, $E=5$ } 
    \label{fig:TOP_mbb_m3_r4}
\end{figure}

\begin{figure}
\centering
  \begin{tabular}{cc}
   \subfigure[]{
     \centering
        \includegraphics[width=38mm]{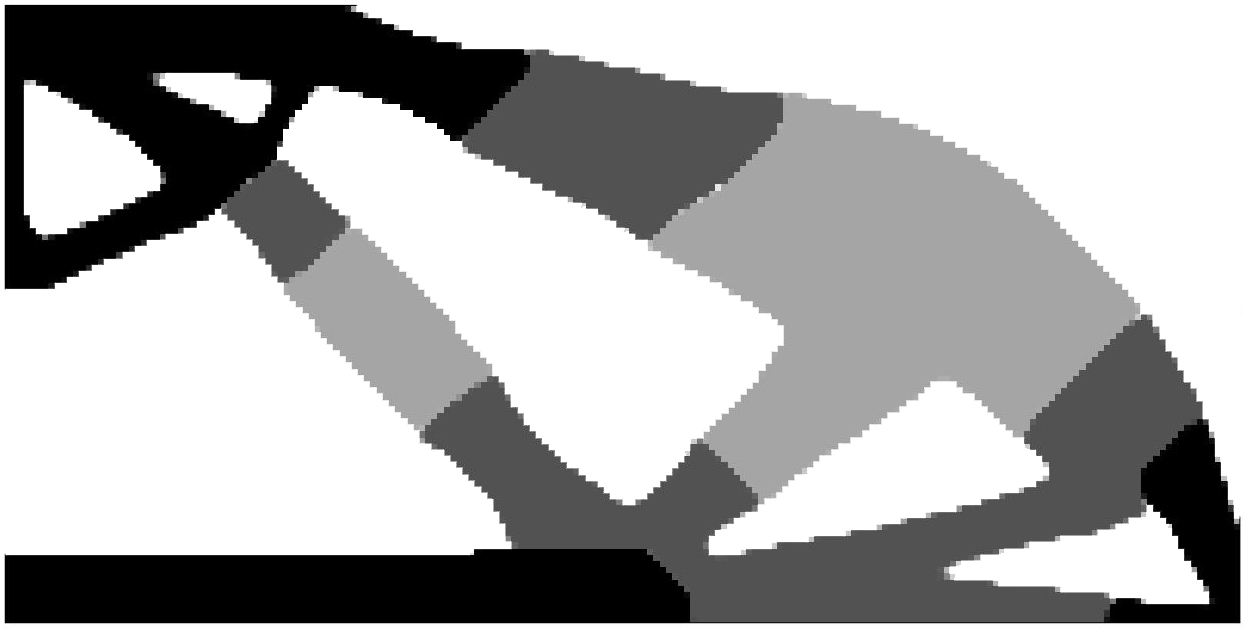}}
    \subfigure[]{
     \centering
        \includegraphics[width=38mm]{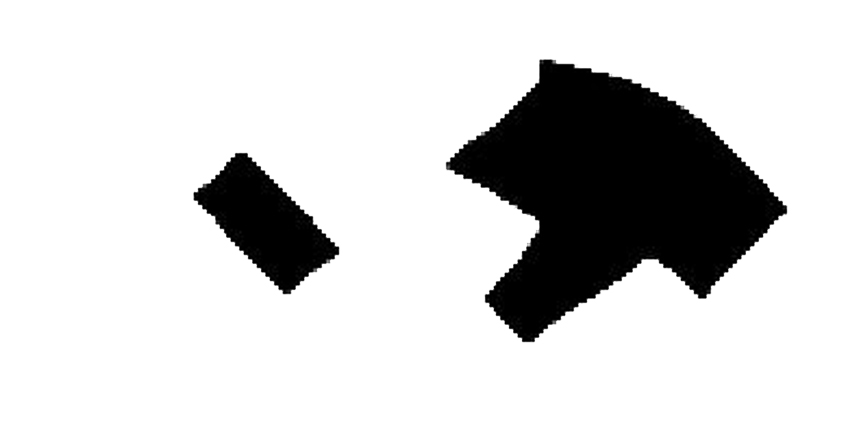}          }\\        
    \subfigure[]{
        \includegraphics[width=38mm]{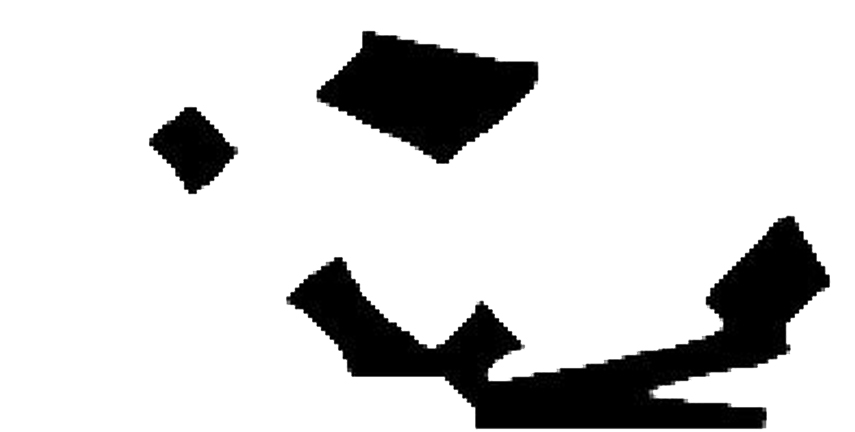}          }
  \subfigure[]{
     \centering
        \includegraphics[width=38mm]{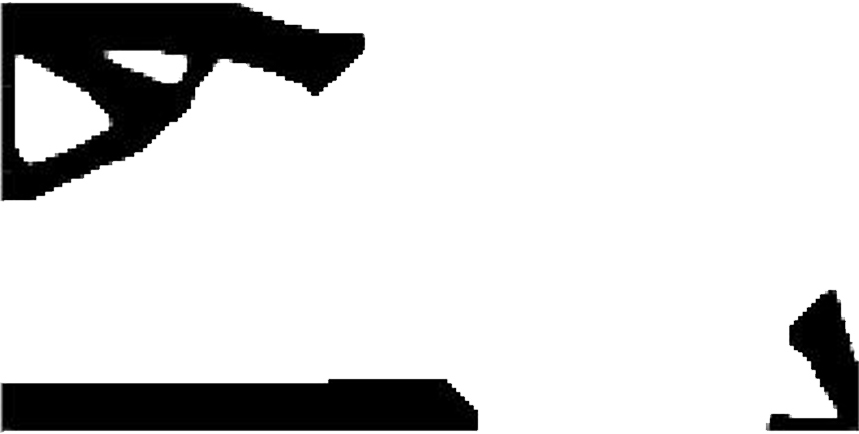}          }\\        
  \end{tabular}        
    \caption{Topology optimization of MBB with 3 materials with $R_{min}=\frac{10r_e}{2\sqrt{3}}$, $V_1=V_2=V_3=\frac{0.5}{3}$, compliance = 0.274 (a) total structure, different colors represent different materials (b) material 1, $E=1$ (c) material 2, $E=2$ (d) material 3, $E=5$ } 
    \label{fig:TOP_mbb_m3_r10}
\end{figure}

We also explored the performance of the proposed method for feature size controlling by changing the value of the parameter $R_{min}$ of the Helmholtz type PDE filter method. The parameter $\beta$ of the projection method is set as the same of the above section. The optimized structures of the proposed method with $R_{min}=\frac{2r_e}{2\sqrt{3}}$, $R_{min}=\frac{4r_e}{2\sqrt{3}}$, $R_{min}=\frac{10r_e}{2\sqrt{3}}$ are shown in Fig.\ref{fig:TOP_mbb_m3_r2}, \ref{fig:TOP_mbb_m3_r4}, \ref{fig:TOP_mbb_m3_r10}, respectively. It also can be found that the optimizer fill strong material near the periphery of the structure where stiffness is needed, and obtain a comparable stiff structure. One interesting thing is found that the proposed method can converge to clear boundaries for each material even with a large filter radius $R_{min}=\frac{10r_e}{2\sqrt{3}}$, which is the same with the simple sensitivity filter based on the average weight of the  radius $R=10r_e$ of the neighbourhood. It is obviously that the mixing at the interfaces resulting from the filter is avoided with the proposed method for all the cases, which is an important contribution of the proposed method. It is easy to manufacture the optimized structure, and would make the simultaneous physical performance for the optimization and manufactured structure.

\subsection{Comparison}

\begin{figure}
\centering
  \begin{tabular}{cc}
    \subfigure[]{
     \centering
        \includegraphics[width=40mm]{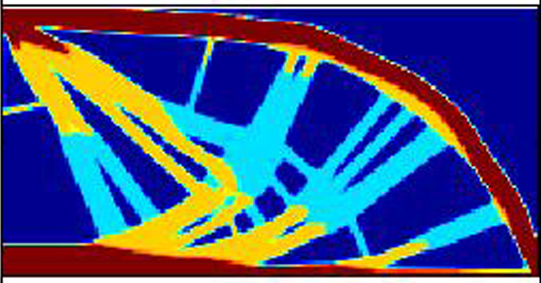}}
    \subfigure[]{
     \centering
        \includegraphics[width=40mm]{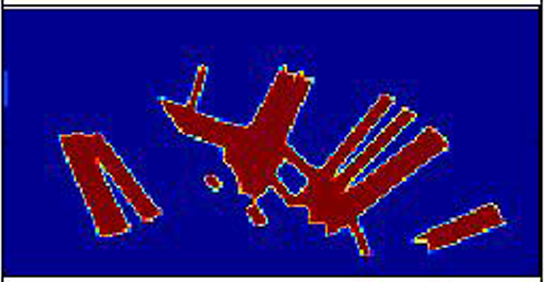}          }\\        
    \subfigure[]{
        \includegraphics[width=40mm]{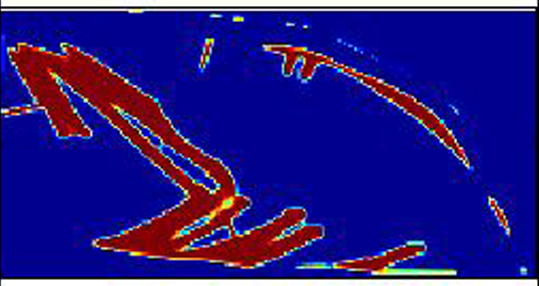}          }
  \subfigure[]{
     \centering
        \includegraphics[width=40mm]{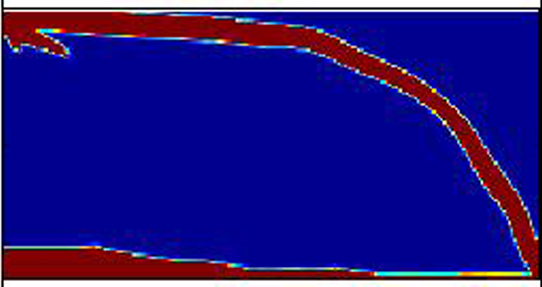}          }\\        
  \end{tabular}        
    \caption{Topology optimization result with the extended SIMP interpolation (a) the optimized layouts of three materials with different colors represent different materials(red is the stiff material $E=5$, cyan is the the compliant material $E=2$, and yellow it is the the material with $E=1$), $V_1=V_2=V_3=\frac{0.5}{3}$, compliance=0.261 (b), (c) and (d) are the details for each material of the optimized structure(red is material, and blue is the void)}
    \label{fig:TOP_SD}
\end{figure}

\begin{figure}
\centering
  \begin{tabular}{cc}
    \subfigure[]{
     \centering
        \includegraphics[width=40mm]{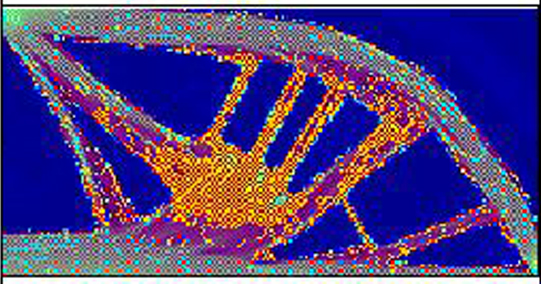}}
    \subfigure[]{
     \centering
        \includegraphics[width=40mm]{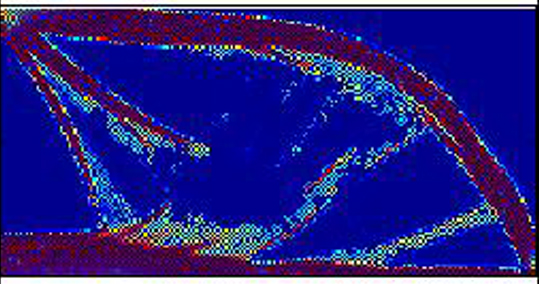}          }\\        
    \subfigure[]{
        \includegraphics[width=40mm]{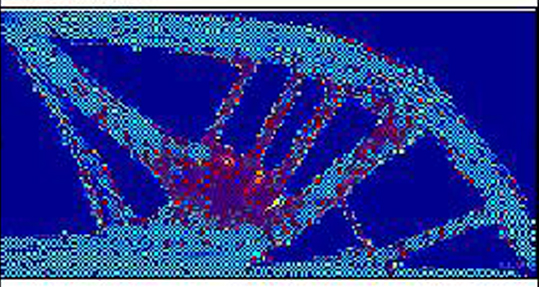}          }
  \subfigure[]{
     \centering
        \includegraphics[width=40mm]{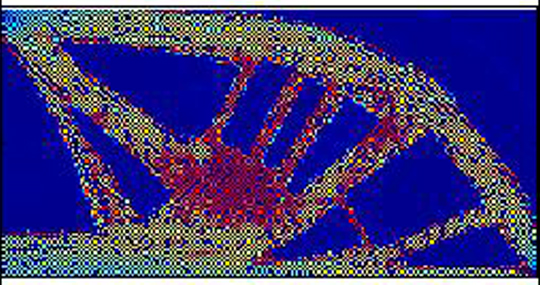}          }\\        
  \end{tabular}        
    \caption{Topology optimization result with the extended SIMP interpolation (a) the optimized layouts of three materials with $E_1 = 5$,$E_2 = 2 $, $E_3 = 1$(the colors represent the mixing of each material), $V_1=V_2=V_3=\frac{0.5}{3}$, compliance=0.268 (b), (c) and (d) are the details for each material of the optimized structure(the colors represent the grey elements of each material)}
    \label{fig:TOP_SI}
\end{figure}

\begin{figure}
\centering
  \begin{tabular}{cc}
    \subfigure[]{
     \centering
        \includegraphics[width=40mm]{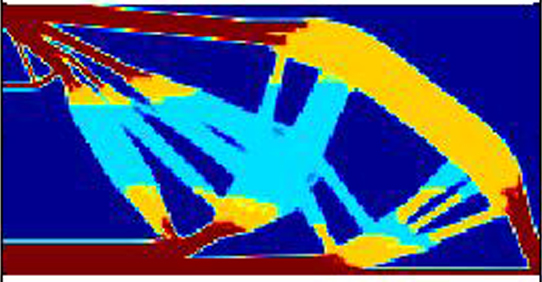}}
    \subfigure[]{
     \centering
        \includegraphics[width=40mm]{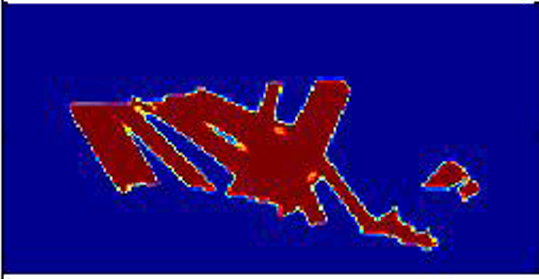}          }\\        
    \subfigure[]{
        \includegraphics[width=40mm]{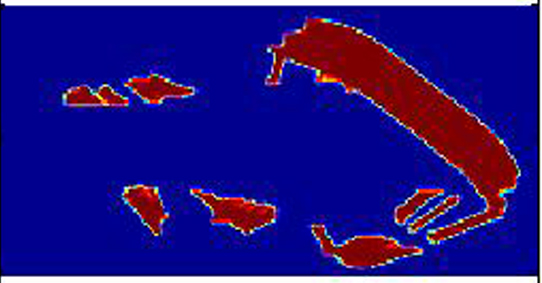}          }
  \subfigure[]{
     \centering
        \includegraphics[width=40mm]{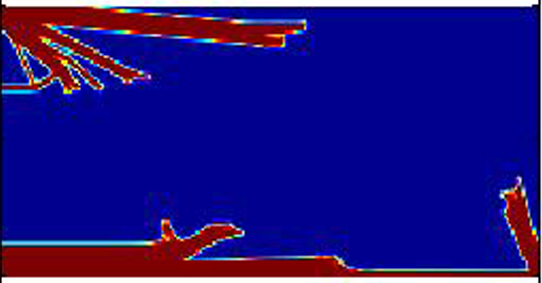}          }\\        
  \end{tabular}        
    \caption{Topology optimization result with the proposed method (a) the optimized layouts of three materials with different colors represent different materials(red is the stiff material $E=5$, cyan is the the compliant material $E=2$, and yellow it is the the material with $E=1$), $V_1=V_2=V_3=\frac{0.5}{3}$, compliance=0.279 (b), (c) and (d) are the details for each material of the optimized structure (red is material, and blue is the void)}
    \label{fig:TOP_pn}
\end{figure}

\begin{figure}
\centering
  \begin{tabular}{cc}
   \subfigure[]{
     \centering
        \includegraphics[width=40mm]{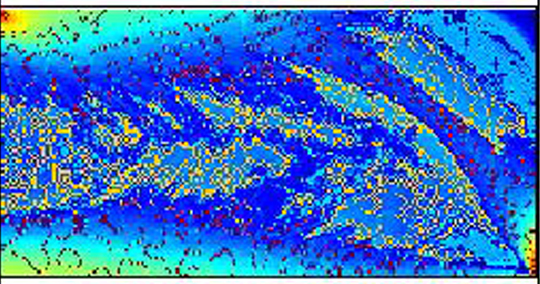}}
    \subfigure[]{
     \centering
        \includegraphics[width=40mm]{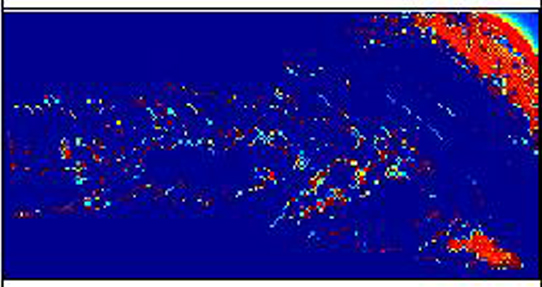}          }\\        
    \subfigure[]{
        \includegraphics[width=40mm]{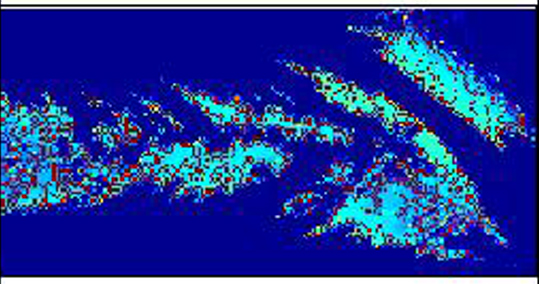}          }
  \subfigure[]{
     \centering
        \includegraphics[width=40mm]{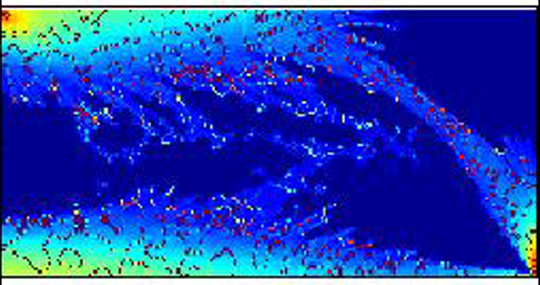}          }\\         
  \end{tabular}        
    \caption{Topology optimization result with the DMO interpolation (a) the optimized layouts of three materials with $E_1 = 1$,$E_2 = 2 $, $E_3 = 5$(the colors represent the mixing of each material),$V_1=V_2=V_3=\frac{0.5}{3}$, compliance=0.170 (b), (c) and (d) are the details for each material of the optimized structure (the colors represent the grey elements of each material)}
    \label{fig:TOP_dmo}
\end{figure}

To illustrate the advantage of the concept of topology optimization for multiple materials with the mapping based interpolation function, we compared the proposed method with both SIMP and DMO based interpolation function via the topology optimization of MMB structure. The conditions of the MBB problem are set as the same of the above section, and the simple sensitivity filter based on the average weight of the radius $R=1.5r_e$ of the neighbourhood is used. All the results and all the finite element codes for these examples can be found in the appendix. As discussed in Fig.~\ref{fig:interpolatedYoungmodulus}, the order of the Young's moduli is important in terms of the convergence of clear 0, 1 result for each element of each material. The decreasing order of Young's modular with SIMP based interpolation function and the DMO based interpolation function(both the increasing and the decreasing order of Young's modular are the same, the only difference is that the optimized structure of each material changing with the order of Young's modular. Thus, we only show the result of the DMO based interpolation function with increasing order of Young's modular.) are difficult to converge to clear 0, 1 result, as shown in Fig.~\ref{fig:TOP_SI}(a) and Fig.~\ref{fig:TOP_dmo}(a), the colors represent the mixing of each material. Fig.~\ref{fig:TOP_SI}(b), (c),(d) and Fig.~\ref{fig:TOP_dmo}(a), (b), (c),(d) represent the grey element of each material. Only, the increasing order of Young's modular with SIMP based interpolation function can obtain a clear layout of each material without few grey elements as shown in Fig.~\ref{fig:TOP_SD}(a), different colors represent different materials(red is the stiff material $E=5$, cyan is the the compliant material $E=2$, and yellow it is the the material with $E=1$). This issue becomes serious when the number of the materials or the phases is increased. As the non-convexity is not observed at the present $p$-norm based interpolation function, the stable convergences can be obtained as shown in Fig.~\ref{fig:TOP_pn}(a), the label is the same with the SIMP based method with increasing order of Young's modular. The compliance is also comparable with that of the SIMP based method as shown in Fig.~\ref{fig:TOP_SD}. Comparing the results of Fig.~\ref{fig:TOP_SD} and Fig.~\ref{fig:TOP_pn}, it can be found out that the boundaries of each material of the optimized structure with the proposed method are much more clearer than the SIMP based method with increasing order of Young's modular, which means that it can be manufactured easily and with fewer lost of the stiffness after manufacturing. Besides, the proposed method can equally and easily initialize each material which may get better optimized structure, and can be easily formulated and implemented when the number of the materials or the phases is increased. 

%%%%%%%%%%%%%%%%%%%%%%%%
\section{Conclusions}
\label{sec4}

This research presents a new mapping based interpolation function for topology optimization with multiple materials. It is common for conventional topology optimization methods of multiple materials to adopt the polynomial functions or the SIMP based interpolation functions. They have been successful for various optimization problems and this research presents a new mapping based interpolation for multiple materials. One of the new features of the present approach is to use the mathematical mappings of the spaces of the design variable. In other words, the material interpolation function is determined by the ratio of the $p$-norm to the 1-norm multiplied by the corresponding design variable. The number of the design variables is the same to the number of the materials of interest. The advantage of the proposed method is that the design variables assigned to each element can be directly used as an indicator whether the corresponding material appears or not, where in the SIMP based interpolation function, the combinations of the design variables should be considered to identify the material used for the volume constraint. Furthermore, the present mapping based interpolation function is different to that of the SIMP based interpolation function. It is tricky to determine which one is better than another. As the design variables indicate the material properties directly, to our experiences, it is relatively easy to find out the initial design variables. It is our opinion that the presented mapping based interpolation function can be used as one of the alternatives of the SIMP based interpolation function. Besides, the proposed method can easily converge to clear 0, 1 result for each material without grey elements, which make it easy to manufacture the optimized structure for multi-materials. The proposed method can also be easily extended and implemented for the topology optimization of plenty materials. 

\section{Acknowledgements}
This work is supported in part by National Natural Science Foundation of China under Grant No.51975589,51875525, The Key Projection of Hunan Province under Grant No.2020GK2096, and the Open Foundation of the State Key Laboratory of Fluid Power and Mechatronic Systems under Grant NO.GZKF-202011 is also gratefully acknowledged.

\section{Ethics declarations}
\subsection{Conflict of interest}
On behalf of all authors, the corresponding author states that there is no conflict of interest.

\subsection{Replication of Results section}
In order to help to understand the content and 
replicate the results. The code for the proposed method, the SIMP and the DMO based interpolation function with both increasing and decreasing order of Young's modular are available as supplementary material for three material. After understanding the approach, the finite element analysis, the sensitivity analysis and the optimization process, it can be easily implemented for topology optimization of structure with more materials. 

%\section*{References}

%\bibliography{Bibliography}

\end{document}